\documentclass[10pt,onecolumn]{revtex4-2}

\usepackage[T1]{fontenc}
\usepackage{amsmath,amsfonts,amssymb,amsthm,bbm}
\usepackage{graphicx}
\usepackage{color}
\usepackage{tipa}

\usepackage[english]{babel}
\usepackage{graphicx}
\usepackage{color}

\usepackage{hyperref}

\begin{document}

\title{$\mu PT$ statistical ensemble: systems with fluctuating energy, particle number, and volume}
\author{Ugo Marzolino}
\affiliation{INFN Trieste Unit, Italy}
\email{ugo.marzolino@ts.infn.it}





\begin{abstract}
Within the theory of statistical ensemble, the so-called $\mu PT$ ensemble describes equilibrium systems that exchange energy, particles, and volume with the surrounding. General, model-independent features of volume and particle number statistics are derived. Non-analytic points of the partition function are discussed in connection with divergent fluctuations and ensemble equivalence. Quantum and classical ideal gases, and a model of Bose gas with mean-field interactions are discussed as examples of the above considerations.
\end{abstract}


\maketitle

\section{Introduction}

Physical systems at equilibrium are studied in statistical mechanics through statistical ensembles. Each ensemble describes a system that exchanges certain physical quantities, typically extensive, with the surrounding. A terminology that is sometimes used in literature \cite{LandauBinder} identifies statistical ensemble with the variables that are held fixed from each statistically conjugated couple $(\mu,N)$, $(P,V)$, and $(\beta,E)$, where $\mu$ is the chemical potential and $N$ is the number of particles, $P$ is the pressure and $V$ is the volume, $\beta=\frac{1}{k_B T}$ is the inverse absolute temperature with $k_B$ being the Boltzmann constant and $E$ is the energy. Accordingly, the microcanonical ensemble describes an isolated system where none of the energy, volume and particles are exchanged, thus called $NVE$. A system in the canonical ensemble exchanges only energy with the surrounding, thus introducing $\beta$ as a parameter of the ensemble which is called $NVT$. The grandcanonical ensemble allows to exchange also particles, with the new parameter $\mu$, and is called $\mu VT$. Finally, the isothermal-isobaric ensemble describes systems exchanging energy and volume with the surrounding, parametrised by $\beta$ and $P$, and is called $NPT$. The $NPT$ ensemble is used in Monte Carlo and molecular dynamics simulations \cite{LandauBinder,Galib2017}.

The above statistical ensembles are related to each other through Legendre transforms \cite{Gallavotti,Attard,Tuckerman}; see also Ref. \cite{Zia2009} for a review on the general framework of Legendre transformations. Moreover, each ensemble is also derived from the maximization of the Shannon entropy with constraints that fix the average of the fluctuating extensive quantities, following the Jaynes' approach \cite{Jaynes1957-1,Jaynes1957-2}. These two constructions are equivalent, resulting in the well-known Boltzmann weight of exponential form. This paper concerns the statistical ensemble, missing in the above picture, which represents a system exchanging energy, particles, and volume with the surrounding, and is parametrised by the intensive variables $\beta$, $\mu$, and $P$, therefore called $\mu PT$ ensemble \cite{Hill,Guggenheim,Campa2018}. The $\mu PT$ ensemble is derived using standard arguments either from the $\mu VT$ ensemble through the Legendre transform with respect to the volume. or from the $NPT$ ensemble through the Legendre transform with respect to the particle number, or again form the maximisation of the Shannon entropy with fixed average energy, particle number and volume.

The $\mu PT$ ensemble is the extension of the $\mu VT$ ensemble when the pressure instead of the volume is fixed, or the extension of the $NPT$ ensemble when the chemical potential instead of the particle number is fixed. The latter conditions are met in several physical and chemical processes, e.g. naturally arise in systems confined within a porous and elastic membranes. Furthermore, the $\mu PT$ ensemble has been studied in small systems \cite{Hill2002,Calabrese2019}, like in nanothermodynamics \cite{Hill2001,Chamberlin2000,Qian2012,Chamberlin2015,Bedeaux2018}, or in systems with long-range interactions \cite{Latella2017,Campa2020}. In these physical systems, the Gibbs-Duhem equation is not supposed to hold, and therefore the three intensive parameters $\mu$, $P$, and $T$ can be independent. In the following, I will investigate general properties of the $\mu PT$ ensemble, and discuss ensemble equivalence in connection with non-analyticities and non-commutativity of Legendre transforms.

\section{General considerations} \label{gen}

Following the Jaynes' approach, the configuration probabilities of the $\mu PT$ ensemble result from the constrained maximisation of the Shannon entropy:

\begin{equation}
\frac{\partial}{\partial p_j}\left(-\sum_{j}p_j\ln p_j-\lambda\left(\sum_jp_j-1\right)-\beta\langle H\rangle-\beta P\langle V\rangle+\mu\langle N\rangle\right)=0,
\end{equation}
where $p_j$ is the probability of the $j$-th configuration with fixed energy, volume and particle number, $H$ is the Hamiltonian, and $\langle\cdot\rangle$ is the average with respect to the probability distribution $\{p_j\}_j$. The configuration probabilities are

\begin{equation}
p_j=\frac{e^{-\beta(E_j+PV_j-\mu N_j)}}{Z_{\mu PT}},
\end{equation}
with the $\mu PT$ partition function

\begin{equation} \label{part.ft}
Z_{\mu PT}=\sum_{\substack{j\in\textnormal{configurations}}}e^{-\beta(E_j+PV_j-\mu N_j)},
\end{equation}
where the sum runs over all configurations at different energy, particle numbers, and volume. According to the order in which the configurations are summed, the $\mu PT$ partition function can be written either as the Legendre transform of the $NPT$ ensemble, i.e.

\begin{equation} \label{NPT.to.muPT}
Z_{\mu PT}=\sum_{N=N_1}^{N_2} e^{\beta\mu N}Z_{NPT},
\end{equation}
or as the Legendre transform of the $\mu VT$ ensemble, i.e.

\begin{equation} \label{muVT.to.muPT}
Z_{\mu PT}=\int_{V_1}^{V_2}\frac{dV}{V_0}e^{-\beta PV}Z_{\mu VT},
\end{equation}
where $V_0$ is a constant with the dimension of a volume in order $Z_{\mu PT}$ to be dimensionless, but does not affect physical quantities.

The logarithms of the partition function of statistical ensembles give thermodynamic potentials. These thermodynamic potentials, summarised in table \ref{thermodyn.pot}, are defined by means of thermal averages of extensive quantities, and can also be expressed as linear homogeneous functions of fixed extensive parameters using the Euler's theorem \cite{Tuckerman}. In the $\mu PT$ ensemble however, all possible Legendre transforms have been performed, such that thermal properties only depend on intensive parameters which gauge the extensive thermal averages.
Without additional statistically conjugated couples, the intensive $\mu PT$ thermodynamic potential is connected to finite-size effects \cite{Hill,Campa2018}.

\begin{table}[htbp!]
\center
\begin{tabular}{c|rcl}
{\bf ensemble} & {\bf thermodynamic potential} & & {\bf extensive parameter dependence} \\
\hline
\hline
$NVE$ &
$S$ & $=$ & $(E+PV-\mu N)/T$ \\
\hline
$NVT$ &
$F=\langle H\rangle-TS$ & $=$ & $-PV+\mu N$ \\
\hline
$\mu VT$ & $\langle H\rangle-\mu\langle N\rangle-TS$ & $=$ & $-PV$ \\
\hline
$NPT$ &
$G=\langle H\rangle+P\langle V\rangle-TS$ & $=$ & $\mu N$ \\
\hline
\end{tabular}
\caption{Thermodynamic potentials}
\label{thermodyn.pot}
\end{table}


\section{Legendre transform of the $\mu VT$ ensemble}
\label{muVT}

The $\mu VT$ partition function is

\begin{equation} \label{ZmuVT}
Z_{\mu VT}=e^{\beta P_c V},
\end{equation}
where $P_c$ is the pressure derived in the $\mu VT$ ensemble.
It is crucial to note that $P_c$ depends only on the free parameters of the $\mu VT$ ensemble, namely the chemical potential $\mu$, the temperature $T$ and the volume $V$, as all other physical quantities are functions of these parameters. Requiring that $P_c$, $\mu$, and $T$, are all intensive quantities, and that the $\mu VT$ thermodynamic potential, i.e. $-P_cV$, is extensive implies that $P_c$ is a non-inceasing function of $V$. In particular, the leading contribution to $P_c$ for large volume does not depend on $V$.

The $\mu PT$ partition function is thus

\begin{equation} \label{legendre.muVT}
Z_{\mu PT}=\int_{V_1}^{V_2}\frac{dV}{V_0}e^{-\beta PV}Z_{\mu VT}=\frac{e^{\beta V_2(P_c-P)}-e^{\beta V_1(P_c-P)}}{\beta V_0(P_c-P)},
\end{equation}
where the volume fluctuates in the interval $[V_1,V_2]$. The thermodynamic potential is

\begin{align}
-\frac{1}{\beta}\ln Z_{\mu PT} & =-V_2(P_c-P)-\frac{1}{\beta}\ln\left(1-e^{-\beta(V_2-V_1)(P_c-P)}\right)+\frac{1}{\beta}\ln(\beta V_0(P_c-P)) \nonumber \\
& =-V_1(P_c-P)-\frac{1}{\beta}\ln\left(1-e^{\beta(V_2-V_1)(P_c-P)}\right)+\frac{1}{\beta}\ln(\beta V_0(P-P_c)),
\end{align}
which becomes, for $\beta(V_2-V_1)(P_c-P)\gg 1$,

\begin{equation}
-\frac{1}{\beta}\ln Z_{\mu PT}=-V_2(P_c-P)+\frac{1}{\beta}\ln(\beta V_0(P_c-P))-\frac{1}{\beta} \, \mathcal{O}\left(e^{-\beta(V_2-V_1)(P_c-P)}\right),
\end{equation}
while, for $\beta(V_2-V_1)(P-P_c)\gg 1$,

\begin{equation}
-\frac{1}{\beta}\ln Z_{\mu PT}=-V_1(P_c-P)+\frac{1}{\beta}\ln(\beta V_0(P_c-P))-\frac{1}{\beta} \, \mathcal{O}\left(e^{-\beta(V_2-V_1)(P-P_c)}\right),
\end{equation}
and, at $P=P_c$,

\begin{equation}
-\frac{1}{\beta}\ln Z_{\mu PT}=-\frac{1}{\beta}\ln\frac{V_2-V_1}{V_0}.
\end{equation}
Notice that $P=P_c$ is a non-analyticity point of the thermodynamic potential when $V_2\to\infty$, which shall be characterised in the following.

\subsection{Volume statistics} \label{vol.stat}

\begin{figure}[htbp]
\centering
\includegraphics[width=0.7\columnwidth]{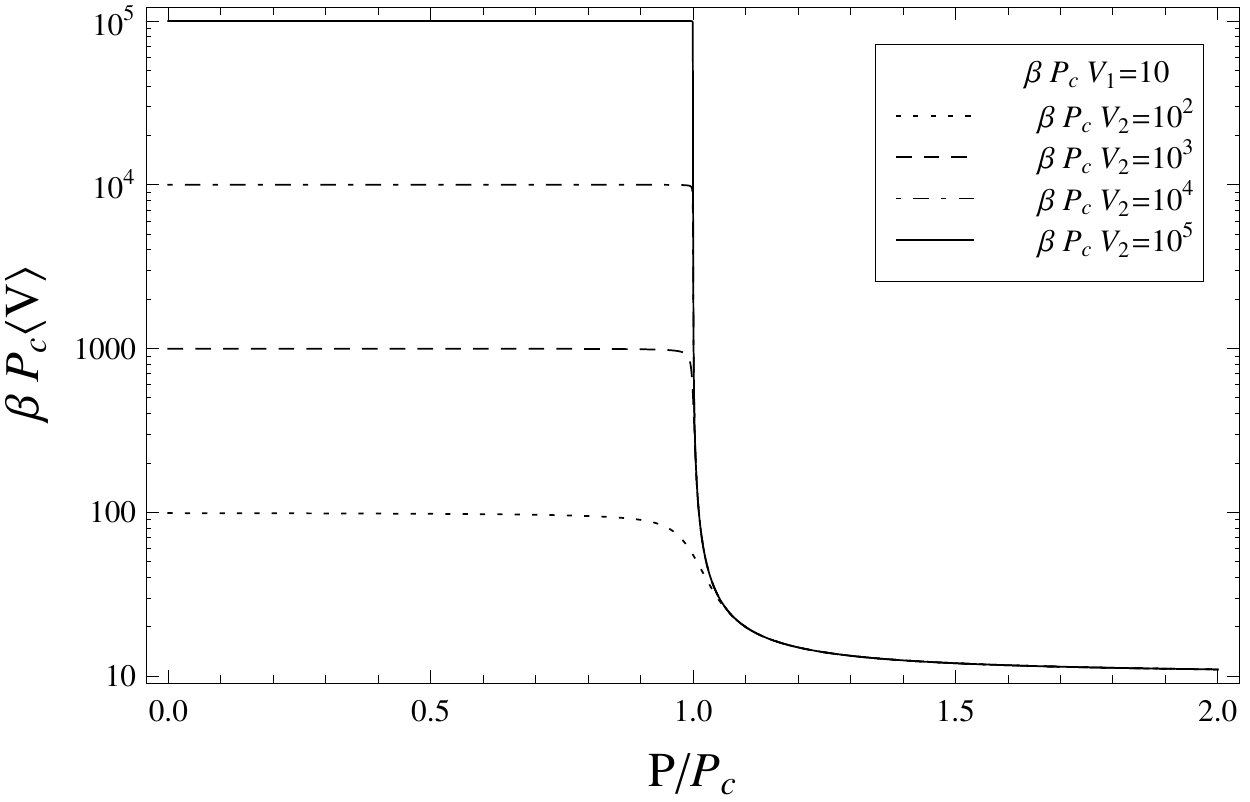}
\caption{Semi-log plot of the rescaled average volume $\beta P_c\langle V\rangle$ as a function of the rescaled pressure $P/P_c$.}
\label{average-volume}
\end{figure}

General features of the volume statistics are derived from pressure derivatives of \eqref{legendre.muVT}. The average volume is

\begin{equation} \label{vol.av}
\langle V\rangle=-\frac{1}{\beta}\frac{\partial}{\partial P}\ln Z_{\mu PT}=\frac{1}{\beta(P-P_c)}+\frac{V_2-V_1 e^{-\beta(V_2-V_1)(P_c-P)}}{1-e^{-\beta(V_2-V_1)(P_c-P)}}
\end{equation}
with the following asymptotic behaviours

\begin{equation}
\langle V\rangle=
\begin{cases}
\displaystyle V_2+\frac{1}{\beta(P-P_c)}+(V_2-V_1)\mathcal{O}\left(e^{-\beta(V_2-V_1)(P_c-P)}\right) & \textnormal{if } \beta(V_2-V_1)(P_c-P)\gg 1 \\
\displaystyle V_1+\frac{1}{\beta(P-P_c)}+(V_2-V_1)\mathcal{O}\left(e^{-\beta(V_2-V_1)(P-P_c)}\right) & \textnormal{if } \beta(V_2-V_1)(P-P_c)\gg 1 \\
\displaystyle \frac{V_2+V_1}{2}+(V_2-V_1)\mathcal{O}\big(\beta(V_2-V_1)|P_c-P|\big) & \textnormal{if } \beta(V_2-V_1)|P_c-P|\ll 1
\end{cases}.
\end{equation}
Therefore, the average volume has a discontinuity when $V_2\to\infty$, as depicted in figure \ref{average-volume}.

\begin{figure}[htbp]
\centering
\includegraphics[width=0.7\columnwidth]{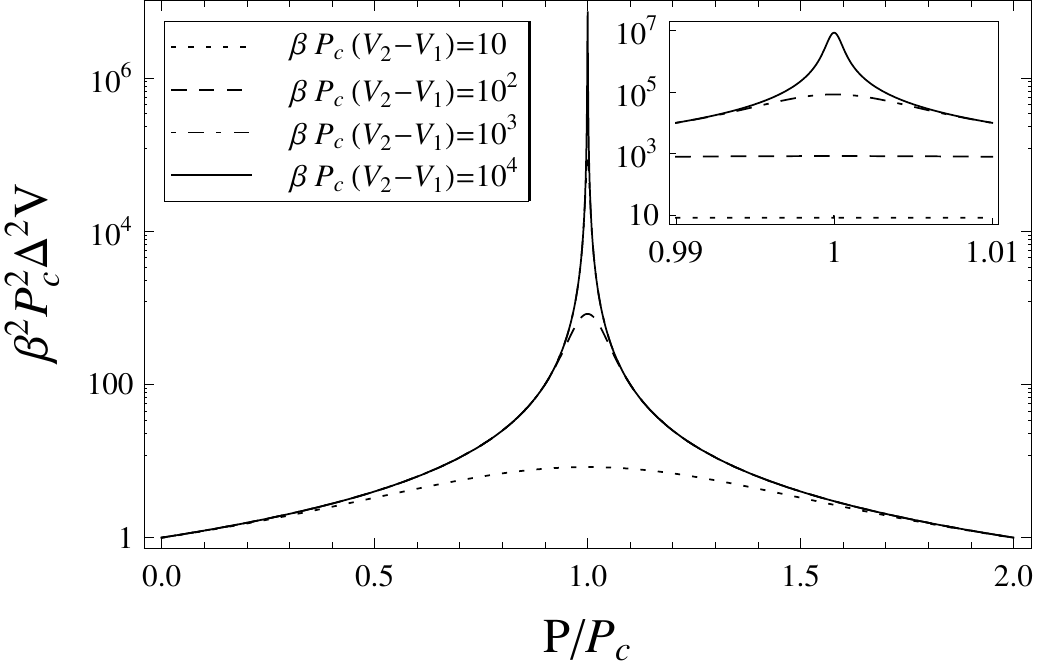}
\caption{Semi-log plot of the rescaled volume fluctuations $\beta^2 P_c^2\Delta^2V$ as a function of the rescaled pressure $P/P_c$.}
\label{fluct-volume}
\end{figure}

The variance of the volume is proportional to the isothermal compressibility $\kappa_T$:

\begin{align}
\label{vol.fluct0}
\Delta^2V & =\frac{1}{\beta^2}\frac{\partial^2}{\partial P^2}\ln Z_{\mu PT}=-\frac{1}{\beta}\frac{\partial}{\partial P}\langle V\rangle=\frac{\kappa_T}{\beta}\langle V\rangle= \nonumber \\
& =\frac{1}{\beta^2(P_c-P)^2}-\frac{(V_2-V_1)^2}{4\sinh^2\left(\frac{\beta}{2}(V_2-V_1)(P_c-P)\right)},
\end{align}
plotted in figure \ref{fluct-volume}, whose asymptotic behaviours are

\begin{equation} \label{vol.fluct1}
\Delta^2V=
\begin{cases}
\displaystyle \frac{1}{\beta^2(P_c-P)^2}-(V_2-V_1)^2\mathcal{O}\left(e^{-\beta(V_2-V_1)|P_c-P|}\right) & \textnormal{if } \beta(V_2-V_1)|P_c-P|\gg 1 \\
\displaystyle (V_2-V_1)^2\left(\frac{1}{12}+\mathcal{O}\big(\beta(V_2-V_1)|P_c-P|\big)^2\right) & \textnormal{if } \beta(V_2-V_1)|P_c-P|\ll 1
\end{cases}.
\end{equation}
Thus, the variance of the volume is superextensive around $P=P_c$ and intensive away from $P=P_c$, as shown in figure \ref{fluct-volume}.

\subsection{Particle number statistics} \label{numb.stat.from.muVT}

The general relation between the average volume and the average particle number is straightforwardly derived:

\begin{equation} \label{av.part.numb}
\langle N\rangle=\frac{1}{\beta}\frac{\partial}{\partial\mu}\ln Z_{\mu PT}=\left(\frac{\partial}{\partial\mu}P_c\right)\langle V\rangle.
\end{equation}
Recalling the definition \eqref{ZmuVT}, the latter is the same equation as in the $\mu VT$ ensemble with the volume, that is a fixed parameter in the grandcanonical ensemble, replaced by the average volume.

The general relation between the particle number and the volume fluctuations is

\begin{equation} \label{part.numb.fluct}
\Delta^2 N=\frac{1}{\beta^2}\frac{\partial^2}{\partial\mu^2}\ln Z_{\mu PT}=\left(\frac{\partial^2}{\partial\mu^2}P_c\right)\frac{\langle V\rangle}{\beta}+\frac{\langle N\rangle^2}{\langle V\rangle^2}\Delta^2 V,
\end{equation}
where the first term in the right-hand-side equals the variance of the particle number in the $\mu VT$ ensemble with the volume replaced by the average volume $\langle V\rangle$.

\subsection{Density statistics}

Since both particle number and volume are fluctuating quantities, also the density $\rho=N/V$ fluctuates. The average density is

\begin{equation} \label{av.density}
\langle\rho\rangle=\frac{1}{Z_{\mu PT}}\int_{V_1}^{V_2}\frac{dV}{V_0}\textnormal{Tr}\left(\frac{N}{V}e^{-\beta(H+PV-\mu N)}\right)=\frac{1}{\beta Z_{\mu PT}}\frac{\partial}{\partial\mu}\int_{V_1}^{V_2}\frac{dV}{V_0 V}\textnormal{Tr}\left(e^{-\beta(H+PV-\mu N)}\right)=\frac{\partial}{\partial\mu}P_c=\frac{\langle N\rangle}{\langle V\rangle},
\end{equation}
and equals the density in the $\mu VT$ ensemble. Equation \eqref{av.density} shows that there is no ambiguity to define the average density as the average of $\frac{N}{V}$ or as the ratio of averages $\frac{\langle N\rangle}{\langle V\rangle}$.

\begin{figure}[htbp]
\centering
\includegraphics[width=0.7\columnwidth]{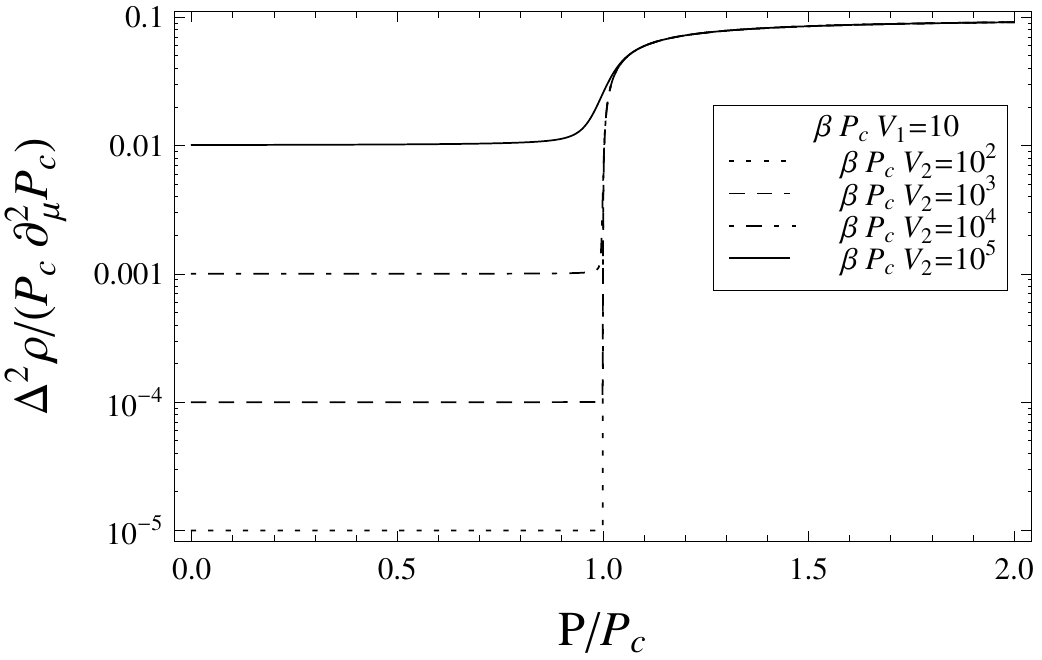}
\caption{Semi-log plot of the rescaled density fluctuations $\frac{\Delta^2\rho}{P_c\partial^2_\mu P_c}$ as a function of the rescaled pressure $P/P_c$.}
\label{fluct-density}
\end{figure}

The variance of the density is

\begin{align}
\Delta^2\rho=\langle\rho^2\rangle-\langle\rho\rangle^2 & =\left(\frac{\partial^2}{\partial\mu^2}P_c\right)\left(P_c-P\right)\frac{\textnormal{Ei}(\beta V_2(P_c-P))-\textnormal{Ei}(\beta V_1(P_c-P))}{e^{\beta V_2(P_c-P)}-e^{\beta V_1(P_c-P)}} \nonumber \\
& =\frac{\Delta^2N-\langle\rho\rangle^2\Delta^2V}{\langle V\rangle}\beta\left(P_c-P\right)\frac{\textnormal{Ei}(\beta V_2(P_c-P))-\textnormal{Ei}(\beta V_1(P_c-P))}{e^{\beta V_2(P_c-P)}-e^{\beta V_1(P_c-P)}},
\label{dens.var}
\end{align}
with $\textnormal{Ei}(x)=-\int_{-x}^\infty dt\frac{e^{-t}}{t}$ being the exponential integral \cite{AbramowitzStegun}, and is sketched in figure \ref{fluct-density}. The asymptotic limits of the density fluctuations are

\begin{equation} \label{dens.var.asymp1}
\Delta^2\rho=
\frac{1}{\beta(V_2-V_1)}\left(\frac{\partial^2}{\partial\mu^2}P_c\right)\left(\ln\frac{V_2}{V_1}+\mathcal{O}\big(\beta V_{1,2}|P_c-P|\big)\right) \qquad \textnormal{if} \qquad \beta V_{1,2}|P_c-P|\ll1,
\end{equation}
recalling the series\cite{BenderOrszag} $\textnormal{Ei}(x)=\gamma+\ln |x|+\sum_{k=1}^\infty\frac{x^k}{k\cdot k!}$ with $\gamma$ the Euler-Mascheroni constant,
while, using instead the asymptotic series \cite{BleisteinHandelsman} $\textnormal{Ei}(x)=\frac{e^x}{x}\sum_{k=0}^{M-1}\frac{k!}{x^k}+\mathcal{O}\big(M!x^{-M}\big)$,

\begin{equation} \label{dens.var.asymp2}
\Delta^2\rho=
\begin{cases}
\displaystyle \frac{1}{\beta V_2}\left(\frac{\partial^2}{\partial\mu^2}P_c\right)\left(1+\mathcal{O}\left(\frac{1}{\beta V_2(P_c-P)}\right)\right) \qquad \textnormal{if} \qquad \beta(V_2-V_1)(P_c-P)\gg1 \\
\displaystyle \frac{1}{\beta V_1}\left(\frac{\partial^2}{\partial\mu^2}P_c\right)\left(1+\mathcal{O}\left(\frac{1}{\beta V_1(P_c-P)}\right)\right) \qquad \textnormal{if} \qquad \beta(V_2-V_1)(P-P_c)\gg1
\end{cases}.
\end{equation}
Therefore, the variance of the density satisfies the so-called shot-noise limit $\Delta\rho=\mathcal{O}\big(1/\langle V\rangle\big)$ for $\beta(V_2-V_1)|P_c-P|\gg1$, and the shot-noise limit with multiplicative logarithmic corrections for $\beta V_{1,2}|P_c-P|\ll1$.

\subsection{Energy statistics}

The average energy is

\begin{equation} \label{av.en.quant}
\langle H\rangle=-\frac{\partial}{\partial\beta}\ln Z_{\mu PT}-P\langle V\rangle+\mu\langle N\rangle=\left(\mu\frac{\partial}{\partial\mu}P_c-\beta\frac{\partial}{\partial\beta}P_c-P_c\right)\langle V\rangle,
\end{equation}
which again equals the expression of the $\mu VT$ ensemble with the volume replaced by the average volume.

Heat capacities at constant volume and at constant pressure are related to derivatives of the mean energy, particle number, and volume:

\begin{align}
C_V= & \left(\frac{dQ}{dT}\right)_{\langle V\rangle}=\left(\frac{d\langle H\rangle}{dT}-\mu\frac{d\langle N\rangle}{dT}\right)_{\langle V\rangle}=k_B\beta^2\left(2\,\frac{\partial}{\partial\beta}P_c+\beta\frac{\partial^2}{\partial\beta^2}P_c\right)\langle V\rangle, \\
C_P= & \left(\frac{dQ}{dT}\right)_P=\left(\frac{d\langle H\rangle}{dT}+P\frac{d\langle V\rangle}{dT}-\mu\frac{d\langle N\rangle}{dT}\right)_P=C_V+k_B\beta^2\left(P-P_c-\beta\frac{\partial P_c}{\partial\beta}\right)^2\Delta^2V.
\end{align}

\section{Legendre transform of the $NPT$ ensemble}
\label{NPT}

The $NPT$ partition function is

\begin{equation} \label{ZNPT}
Z_{NPT}=e^{-\beta\mu_c N},
\end{equation}
where $\mu_c$ is the chemical potential derived in the $NPT$ ensemble. In analogy to the discussion after equation \eqref{ZmuVT}, note that $\mu_c$ depends only on the free parameters of the $NPT$ ensemble, namely the pressure $P$, the temperature $T$ and the particle number $N$. Given that $\mu_c$, $P$, and $T$, are intensive, and that the $NPT$ thermodynamic potential, i.e. $\mu_cN$, is extensive, $\mu_c$ is a non-inceasing function of $N$. In particular, the leading contribution to $\mu_c$ for large volume does not depend on $N$.

The $\mu PT$ partition function is

\begin{equation} \label{legendre.NPT}
Z_{\mu PT}=\sum_{N_1}^{N_2}e^{\beta\mu N}Z_{NPT}=\frac{e^{\beta N_1(\mu-\mu_c)}-e^{\beta(N_2+1)(\mu-\mu_c)}}{1-e^{\beta(\mu-\mu_c)}},
\end{equation}
where the number of particles fluctuates in the interval $[N_1,N_2]$. The thermodynamic potential is

\begin{align}
-\frac{1}{\beta}\ln Z_{\mu PT} & =-(N_2+1)(\mu-\mu_c)-\frac{1}{\beta}\ln\left(1-e^{-\beta(N_2-N_1+1)(\mu-\mu_c)}\right)+\frac{1}{\beta}\ln\left(e^{\beta(\mu-\mu_c)}-1\right) \nonumber \\
& =-N_1(\mu-\mu_c)-\frac{1}{\beta}\ln\left(1-e^{\beta(N_2-N_1+1)(\mu-\mu_c)}\right)+\frac{1}{\beta}\ln\left(1-e^{\beta(\mu-\mu_c)}\right),
\end{align}
which becomes, for $\beta(N_2-N_1+1)(\mu-\mu_c)\gg 1$,

\begin{equation}
-\frac{1}{\beta}\ln Z_{\mu PT}=-(N_2+1)(\mu-\mu_c)+\frac{1}{\beta}\ln\left(e^{\beta(\mu-\mu_c)}-1\right)+\frac{1}{\beta} \, \mathcal{O}\left(e^{-\beta(N_2-N_1+1)(\mu-\mu_c)}\right),
\end{equation}
while, for $\beta(N_2-N_1+1)(\mu_c-\mu)\gg 1$,

\begin{equation}
-\frac{1}{\beta}\ln Z_{\mu PT}=-N_1(\mu-\mu_c)+\frac{1}{\beta}\ln\left(1-e^{\beta(\mu-\mu_c)}\right)+\frac{1}{\beta} \, \mathcal{O}\left(e^{-\beta(N_2-N_1+1)(\mu_c-\mu)}\right),
\end{equation}
and, at $\mu=\mu_c$,

\begin{equation}
-\frac{1}{\beta}\ln Z_{\mu PT}=-\frac{1}{\beta}\ln(N_2-N_1+1).
\end{equation}
Therefore, $\mu=\mu_c$ is a non-analyticity point of the thermodynamic potential when $N_2\to\infty$.

\subsection{Particle number statistics} \label{part.numb.stat}

\begin{figure}[htbp]
\centering
\includegraphics[width=0.7\columnwidth]{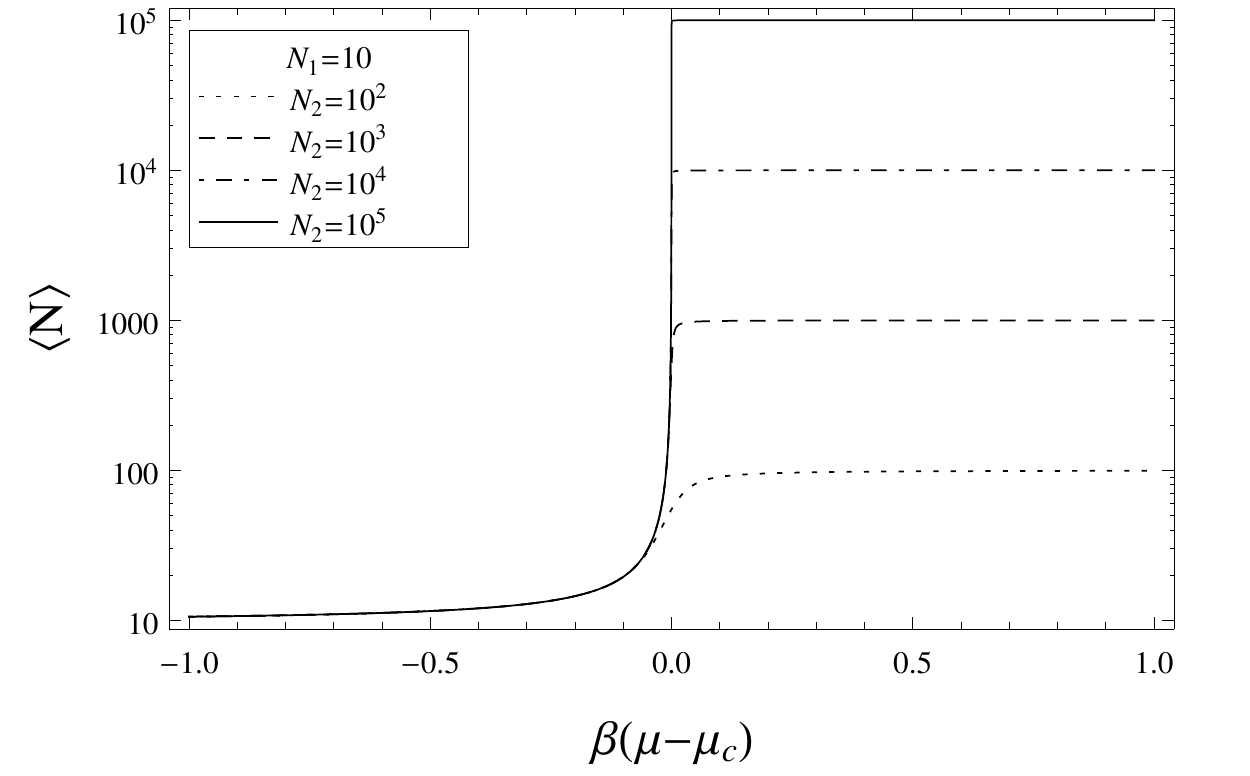}
\caption{Semi-log plot of the average number of particles $\langle N\rangle$ as a function of the rescaled chemical potential $\beta(\mu-\mu_c)$.}
\label{average-part-numb}
\end{figure}

General features of the particle number statistics are derived from derivatives of \eqref{legendre.NPT} with respect to the chemical potential. The average number of particles is 

\begin{equation} \label{gen.av.part.numb}
\langle N\rangle=\frac{1}{\beta}\frac{\partial}{\partial\mu}\ln Z_{\mu PT}=\frac{1}{e^{-\beta(\mu-\mu_c)}-1}+\frac{N_2+1-N_1 e^{-\beta(N_2-N_1+1)(\mu-\mu_c)}}{1-e^{-\beta(N_2-N_1+1)(\mu-\mu_c)}},
\end{equation}
with the following asymptotic behaviours

\begin{equation}
\langle N\rangle=
\begin{cases}
\displaystyle N_2+\frac{1}{1-e^{\beta(\mu-\mu_c)}}+(N_2-N_1+1)\mathcal{O}\left(e^{-\beta(N_2-N_1)(\mu-\mu_c)}\right) & \textnormal{if } \beta(N_2-N_1)(\mu-\mu_c)\gg 1 \\
\displaystyle N_1+\frac{1}{e^{\beta(\mu_c-\mu)}-1}+(N_2-N_1+1)\mathcal{O}\left(e^{-\beta(N_2-N_1)(\mu_c-\mu)}\right) & \textnormal{if } \beta(N_2-N_1)(\mu_c-\mu)\gg 1 \\
\displaystyle \frac{N_2+N_1}{2}+(N_2-N_1-2)\mathcal{O}\big(\beta(N_2-N_1)|\mu-\mu_c|\big) & \textnormal{if } \beta(N_2-N_1)|\mu-\mu_c|\ll 1
\end{cases}.
\end{equation}
As for the average volume, the number of particles has a discontinuity when $N_2\to\infty$, as depicted in figure \ref{average-part-numb}.

\begin{figure}[htbp]
\centering
\includegraphics[width=0.7\columnwidth]{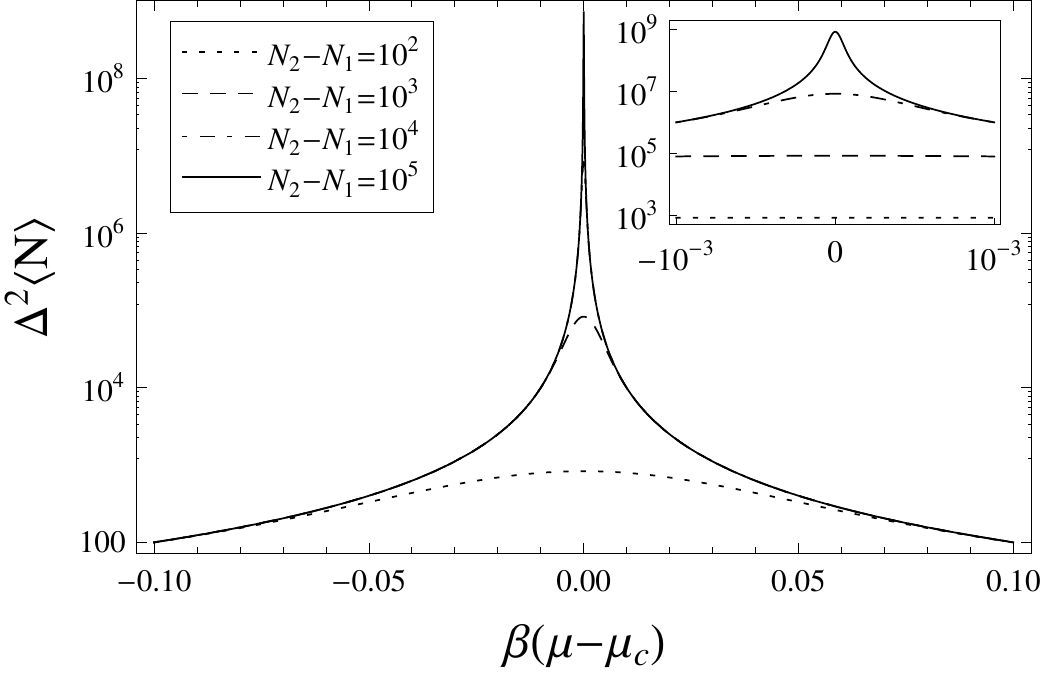}
\caption{Semi-log plot of the particle number fluctuations $\Delta^2N$ as a function of the rescaled chemical potential $\beta(\mu-\mu_c)$.}
\label{fluct-part-numb}
\end{figure}

The variance of the particle number is

\begin{align} \label{gen.var.part.numb}
\Delta^2N & =\frac{1}{\beta^2}\frac{\partial^2}{\partial\mu^2}\ln Z_{\mu PT}=\frac{1}{\beta}\frac{\partial}{\partial\mu}\langle N\rangle= \nonumber \\
& =\frac{1}{4\sinh^2\left(\frac{\beta}{2}(\mu-\mu_c)\right)}-\frac{(N_2-N_1+1)^2}{4\sinh^2\left(\frac{\beta}{2}(N_2-N_1+1)(\mu-\mu_c)\right)},
\end{align}
plotted in figure \ref{fluct-part-numb}, whose asymptotic behaviours are

\begin{equation} \label{gen.var.part.numb1}
\Delta^2N=
\begin{cases}
\displaystyle \frac{1}{4\sinh^2\left(\frac{\beta}{2}(\mu-\mu_c)\right)}-(N_2-N_1+1)^2\mathcal{O}\left(e^{-\beta(N_2-N_1)|\mu-\mu_c|}\right) & \textnormal{if } \beta(N_2-N_1)|\mu-\mu_c|\gg 1 \\
\displaystyle (N_2-N_1)(N_2-N_1+2)\left(\frac{1}{12}+
\mathcal{O}\big(\beta(N_2-N_1)|\mu-\mu_c|\big)^2\right) & \textnormal{if } \beta(N_2-N_1)|\mu-\mu_c|\ll 1
\end{cases}.
\end{equation}
Thus, the variance of the particle number is superextensive around $\mu=\mu_c$ and intensive away from $\mu=\mu_c$, as shown in figure \ref{fluct-part-numb}.

\subsection{Volume statistics}

The general relation between the average particle number and the average volume is again straightforwardly derived:

\begin{equation} \label{av.vol}
\langle V\rangle=-\frac{1}{\beta}\frac{\partial}{\partial P}\ln Z_{\mu PT}=\left(\frac{\partial}{\partial P}\mu_c\right)\langle N\rangle.
\end{equation}
Recalling the definition \eqref{ZNPT}, the latter is the same equation as in the $NPT$ ensemble with the particle number, that is a fixed parameter in the isothermal-isobaric ensemble, replaced by the average particle number.

The general relation between the particle number and the volume fluctuations is

\begin{equation} \label{vol.fluct}
\Delta^2 V=\frac{1}{\beta^2}\frac{\partial^2}{\partial P^2}\ln Z_{\mu PT}=-\frac{1}{\beta}\frac{\partial}{\partial P}\langle V\rangle=\frac{\kappa_T}{\beta}\langle V\rangle=-\left(\frac{\partial^2}{\partial P^2}\mu_c\right)\frac{\langle N\rangle}{\beta}+\frac{\langle V\rangle^2}{\langle N\rangle^2}\Delta^2 N,
\end{equation}
where the first term of equation \eqref{vol.fluct} equals the variance of the volume in the $NPT$ ensemble with the average particle number replaced by the average particle number $\langle N\rangle$.

\subsection{Energy statistics}

The average energy is

\begin{equation} \label{av.en.quant2}
\langle H\rangle=-\frac{\partial}{\partial\beta}\ln Z_{\mu PT}-P\langle V\rangle+\mu\langle N\rangle=\left(\mu_c+\beta\frac{\partial}{\partial\beta}\,\mu_c-P\frac{\partial}{\partial P}\,\mu_c\right)\langle N\rangle,
\end{equation}
which again equals the expression of the $NPT$ ensemble with the particle number replaced by the average particle number.

From the derivative of the averages of energy, volume and particle number, heat capacities at constant volume and at constant pressure are computed:

\begin{align}
\label{CV.legendreNPT}
C_V= & \left(\frac{dQ}{dT}\right)_{\langle V\rangle}=\left(\frac{d\langle H\rangle}{dT}-\mu\frac{d\langle N\rangle}{dT}\right)_{\langle V\rangle}= \nonumber \\
= & k_B\beta^2\left(\mu_c-\mu+\beta\frac{\partial\mu_c}{\partial\beta}\right)\frac{\partial^2\mu_c}{\partial\beta\partial P}\left(\frac{\partial\mu_c}{\partial P}\right)^{-1}\langle N\rangle
-k_B\beta^2\left(2\,\frac{\partial\mu_c}{\partial\beta}+\beta\frac{\partial^2\mu_c}{\partial\beta^2}\right)\langle N\rangle, \\
\label{CP.legendreNPT}
C_P= & \left(\frac{dQ}{dT}\right)_P=\left(\frac{d\langle H\rangle}{dT}+P\frac{d\langle V\rangle}{dT}-\mu\frac{d\langle N\rangle}{dT}\right)_P= \nonumber \\
= & -k_B\beta^2\left(2\,\frac{\partial\mu_c}{\partial\beta}+\beta\frac{\partial^2\mu_c}{\partial\beta^2}\right)\langle N\rangle
+k_B\beta^2\left(\mu_c-\mu+\beta\frac{\partial\mu_c}{\partial\beta}\right)^2\Delta^2N
= \nonumber \\
= & C_V+k_B\beta^2\left(\mu_c-\mu+\beta\frac{\partial\mu_c}{\partial\beta}\right)^2\Delta^2N
-k_B\beta^2\left(\mu_c-\mu+\beta\frac{\partial\mu_c}{\partial\beta}\right)\frac{\partial^2\mu_c}{\partial\beta\partial P}\left(\frac{\partial\mu_c}{\partial P}\right)^{-1}\langle N\rangle.
\end{align}

\section{Non-analyticity} \label{non-analyticity}

The average volume computed from the Legendre transform of the $\mu VT$ ensemble (see equation \eqref{vol.av} and figure \ref{average-volume}) and the average number of particles computed from the Legrndre transform of the $NPT$ ensemble (see equation \eqref{gen.av.part.numb} figure \ref{average-part-numb}) resemble step functions. When $V_2$ ($N_2$) diverges, the average volume (average number of particles) becomes discontinuous at the critical pressure $P_c$ (critical chemical potential $\mu_c$), with a divergent plateau at small pressures $P<P_c$ (large chemical potentials $\mu>\mu_c$). Such discontinuous behaviours resemble a first-order phase transition: e.g. the function $\langle V\rangle(P)$, or its inverse $P(\langle V\rangle)$, is qualitatively similar to the gas-liquid phase transition, and a similar behaviour for $\langle N\rangle(\mu)$ and $\mu(\langle N\rangle)$. The critical point is also characterised by variances of the volume and of the particle number diverging, respectively, as $(V_2-V_1)^2/12$ at $P=P_c$ (see equation \eqref{vol.fluct1} and figure \ref{fluct-volume}) and as $(N_2-N_1)^2/12$ at $\mu=\mu_c$ (see equation \eqref{gen.var.part.numb1} and figure \ref{fluct-part-numb}) for infinitely large $V_2$ and $N_2$, while they are intensive away from the critical values $P=P_c$ and $\mu=\mu_c$.

Fluctuations of the volume are related to fluctuations of the pressure, while fluctuations of the particle number are related to fluctuations of the chemical potential, through uncertainty relations that hold for statistically conjugated variables \cite{Gilmore1985,Falcioni2011,Davis2012,Hiura2018}. Even though pressure and chemical potential are fixed parameters, they can be seen as external fields subject to noise, e.g. in the experimental preparation or in the thermalisation process. Within estimation theory, the variance of pressure and chemical potential estimations, $\delta^2P$ and $\delta^2\mu$, are related to volume and particle number fluctuations by means of the Cram\'er-Rao bound \cite{Cramer,Helstrom,Holevo}

\begin{equation} \label{CRB}
\beta^2\Delta^2V\delta^2P\geqslant\frac{1}{M}, \qquad \beta^2\Delta^2N\delta^2\mu\geqslant\frac{1}{M},
\end{equation}
where $M$ is the number of measurements. Similar relations can also be derived within the mathematical theory of Legendre transforms \cite{Zia2009}. The Cram\'er-Rao bound is formulated using the Fisher information, which is a metric of states (or probability distributions) when they differ by an infinitesimally small parameter change. The Fisher information of the $\mu PT$ esemble equals $\beta^2\Delta^2V$ when pressure is changed and $\beta^2\Delta^2N$ when chemical potential is changed.
Indeed, the estimation of intensive parameters of equilibrium ensembles are related to variances of statistically conjugated extensive variables: $\beta^2\Delta^2X\delta^2\xi\geqslant 1/M$, e.g. with $(\xi,X)=(P,V)$ or $(\xi,X)=(\mu,N)$, and $\Delta^2H\delta^2\beta\geqslant 1/M$.
The connection between the Cram\'er-Rao bound, thermodynamic state geometry, and susceptibilities is discussed in Refs. \cite{Weinhold1975,
Salamon1984,
Diosi1984,
Nulton1985,
Janyszek1986,Janyszek1986-2,
Janyszek1989-2,
Ruppeiner1995,
Brody1995,Dolan1998,Janke2002,Janke2003,
Brody2003,
You2007,
Zanardi2007,Zanardi2008,
Paunkovic2008,Quan2009,Gu2010,Prokopenko2011,
Marzolino2013,Marzolino2015,Braun2018}.
For thermodynamic states away from critical points, variances of extensive variables $X$ are extensive,
implying fluctuations $\Delta X/\langle X\rangle$ vanishing as the inverse of the square root of the system size; the same scaling, known as \emph{shot-noise limit} in metrology, holds for the sensitivity $\delta\xi$ of the intensive parameters $\xi$.

Close to the critical pressure $P=P_c$ and for large $V_2$, the pressure is very close to the pressure in the
$\mu VT$ ensemble with a superextensive volume variance $\Delta^2V=\mathcal{O}\big(\langle V\rangle^2\big)$.
This implies low, i.e. sub-shot-noise, uncertainty for the pressure, $\delta P=\mathcal{O}\big(1/\langle V\rangle\big)$.
On the other hand, away from the critical pressure and for large $V_2$, the variance of the volume is intensive, implying a large variance for the pressure $\delta P=\mathcal{O}(1)$,
and sub-shot-noise scaling for the relative error of the volume $\Delta V/\langle V\rangle=\mathcal{O}\big(1/\langle V\rangle\big)$.
Moreover, when the domain of the volume integration is $[V_1,V_2]\to[0,\infty)$, the leftmost plateau in figure \ref{average-volume} goes to infinity, and the rightmost one assumes an intensive value.
Therefore, the inverse function $P\big(\langle V\rangle\big)$ is almost constant, i.e. equals $P_c$ up to small deviations as discussed above, except for intensive average volume which is not very
relevant for thermodynamic states. In this sense, the pressure in the $\mu PT$ ensemble agrees with that in the $\mu VT$ ensemble, namely $P_c$ which is determined by $\beta$, $\mu$, and $V$.

A completely similar interpretation holds for the variance of the particle number and to that of the chemical potential.
The chemical potential is very close to its value in the $NPT$ ensemble with a superextensive particle number variance $\Delta^2N=\mathcal{O}\big(\langle N\rangle^2\big)$, close to the critical chemical potential $\mu=\mu_c$ and for large $N_2$.
Therefore, the uncertainty for the chemical potential, $\delta\mu=\mathcal{O}\big(1/\langle N\rangle\big)$, obeys sub-shot-noise scaling.
On the other hand, the intensivity of the particle number variance, away from the critical chemical potential and for large $N_2$, implies a large variance for the chemical potential, $\delta\mu=\mathcal{O}(1)$,
but sub-shot-noise limited relative error for the number of particles, $\Delta N/\langle N\rangle=\mathcal{O}\big(1/\langle N\rangle\big)$.
Furthermore, when the domain of the particle number summation is $[N_1,N_2]\to[0,\infty]$, the rightmost plateau in figure \ref{average-part-numb} goes to infinity, and the leftmost one assumes an intensive value.
Therefore, the inverse function $\mu\big(\langle N\rangle\big)$ almost coincides with $\mu_c$, with small deviations $\delta\mu$, except for intensive average number of particles which is not very
relevant for thermodynamic states. In this sense, the chemical potential in the $\mu PT$ ensemble agrees with that in the $NPT$ ensemble, namely $\mu_c$ which is determined by $\beta$, $P$, and $V$.

\section{Ensemble equivalence and non-commutativity of Legendre transforms}
\label{non-comm}

The thermodynamic equivalence between the $\mu PT$ and the $\mu VT$ ensembles require, in particular, that the two ensembles predict the same pressure, namely $P=P_c$. Similarly, thermodynamic equivalence between the $\mu PT$ and the $NPT$ ensembles require $\mu=\mu_c$. Relations $P=P_c$ and $\mu=\mu_c$ are manifestations of the Gibbs-Duhem equation. The need for relations among intensive quantities $\beta$, $\mu$ and $P$
can be derived also from requiring equivalence of other thermal quantities when computed from the Legendre transform of the $\mu VT$ and that of the $NPT$ ensembles.

Consider the density

\begin{equation} \label{consistency}
\frac{\langle N\rangle}{\langle V\rangle}\overset{\substack{\textnormal{from}\\ \textnormal{\eqref{av.part.numb}}}}{=}\frac{\partial P_c}{\partial\mu}\overset{\substack{\textnormal{from}\\\textnormal{\eqref{av.vol}}}}{=}\left(\frac{\partial\mu_c}{\partial P}\right)^{-1}.
\end{equation}
Since $P_c$ is derived from the $\mu VT$ ensemble, it depends only on $\beta$ and $\mu$ and not on $P$, and therefore the same holds for $\frac{\partial P_C}{\partial\mu}$. Analogously, $\mu$ and $\frac{\partial\mu_c}{\partial P}$, being derived using the $NPT$ ensemble, are functions of $\beta$ and $P$ but not of $\mu$. In other words, the left-hand-side of the last equality in \eqref{consistency} is a function only of $\beta$ and $\mu$, while the right-hand-side is a function only of $\beta$ and $P$. Thus, the only possibility to fulfil equation \eqref{consistency} for any value of intensive quantities is that $\frac{\partial P_C}{\partial\mu}$ does not depend on $\mu$ and that $\frac{\partial\mu_c}{\partial P}$ does not depend on $P$. These conditions are so restrictive that are not met in several statistical models, like those studied below. It follows that intensive quantities $\beta$, $\mu$ and $P$ cannot be completely independent, but are constrained by the relation \eqref{consistency}.

Compare now fluctuations \eqref{vol.fluct0} and \eqref{part.numb.fluct}, derived from the Legendre transform of the $\mu VT$ ensemble, with fluctuations \eqref{gen.var.part.numb} and \eqref{vol.fluct}, derived from the Legendre transform of the $NPT$ ensemble.
The fluctuation term $\Delta^2 V$ of the particle number variance \eqref{part.numb.fluct} is superextensive close to the critical point $P=P_c$ and intensive otherwise,
whereas the contribution $\langle V\rangle$ is extensive for $P\leqslant P_c$. Therefore, the particle number variance \eqref{part.numb.fluct}, $\Delta^2 N$, is extensive for $P\leqslant P_c$.
This is in contradiction with the particle number variance \eqref{gen.var.part.numb}, derived from the Legendre transform of the $NPT$ ensemble, which is superextensive close to $\mu=\mu_c$ and intensive otherwise.

Similarly, the term $\Delta^2 N$ of the volume variance \eqref{vol.fluct} is intensive away from the critical point $\mu=\mu_c$,
whereas the contribution $\langle N\rangle$ is extensive for $\mu\geqslant\mu_c$. Therefore, the variance \eqref{vol.fluct}, $\Delta^2 V$, is extensive for
$\mu\geqslant\mu_c$.
On the other hand, the volume variance \eqref{vol.fluct0}, derived from the Legendre transform of the $\mu VT$ ensemble, is superextensive close to $P=P_c$ and intensive otherwise. In conclusion, fluctuations derived from the Legendre transform of the $\mu VT$ ensemble agree with those derived from the Legendre transform of the $NPT$ ensemble only if $P=P_c$ and $\mu=\mu_c$.

The reason for these apparent contradictions is that the Legendre transforms with respect to the volume and to the particle number, both needed to derive the $\mu PT$ ensemble from the $NVT$ ensemble, do not commute. Furthermore, the relations between thermodynamic potentials, $-P_cV$ and $\mu_cN$, with partition functions in equations \eqref{ZmuVT} and \eqref{ZNPT}, respectively, are the leading orders for large system size \cite{Gallavotti}. Therefore,
the order of the Legendre transforms could be dictated by the relative scaling between the volume range and the particle number range, which depends on the specific physical model or on experimental conditions.
A more technical comparison of the $\mu PT$ with other ensembles is provided in Appendix \ref{app}.
The ensemble equivalence can also be checked in specific models, as shall be discussed in the next sections.

\section{Quantum ideal homogeneous gases} \label{quantum.ideal}

The above general results can be exemplified by specific models, e.g. the quantum ideal homogeneous gas in $d$ dimensions. The Hamiltonian is $H=\sum_{j=1}^N\frac{{\bf p}_j^2}{2m}$, with ${\bf p}_j$ the momentum of the $j$-th particle. The pressure in the $\mu VT$ ensemble is

\begin{equation} \label{crit.press.hom.gas}
P_c=\frac{\pm\textnormal{Li}_{\frac{d}{2}+1}(\pm e^{\beta\mu})}{\beta \, \lambda_T^d},
\end{equation}
where $\lambda_T=\sqrt{2\pi h^2\beta/m}$ is the thermal wavelength, $\textnormal{Li}_s(\cdot)$ is the polylogarithm \cite{Wood1992} of order $s$, and the upper (lower) sign holds for Bosons (Fermions).
Using this expression for $P_c$ specifies thermal quantities of the $\mu PT$ ensemble derived as the Legedre transform of the $\mu VT$ ensemble (see section \ref{muVT}). Recall that the derivative of the polylogarithm satisfies $x\frac{\partial}{\partial x}\textnormal{Li}_s(x)=\textnormal{Li}_{s-1}(x)$.

\subsection{Thermal averages and fluctuations}

The relation between the average number of particles and the average volume is

\begin{equation} \label{av.part.numb.quant}
\langle N\rangle=\pm\frac{\langle V\rangle}{\lambda_T^d}\textnormal{Li}_{\frac{d}{2}}(\pm e^{\beta\mu}),
\end{equation}
which is, as the general case \eqref{av.part.numb}, the same equation as in the $\mu VT$ ensemble with the fixed volume replaced by the average volume. In particular, equation \eqref{av.part.numb.quant} for the Bose gas implies that the critical temperature for the Bose-Einstein condensation in the $\mu PT$ ensemble is the same as in the $\mu VT$ one. In fact, the critical temperature is defined by the particle number reaching its an upper bound in the continuum spectrum limit. Thus, when the actual particle number exceeds that bound, low-lying energy levels have macroscopic, and indeed singular in the continuum spectrum limit, occupations.

The average energy is

\begin{equation} \label{av.en.quant.id}
\langle H\rangle=\pm\frac{d\langle V\rangle}{2\beta\lambda_T^d}\textnormal{Li}_{\frac{d}{2}+1}(\pm e^{\beta\mu})=\frac{d}{2}\,P_c\langle V\rangle.
\end{equation}

The relation between volume and particle number fluctuations is

\begin{equation} \label{part.numb.fluct.hom.gas}
\Delta^2 N=\pm\frac{\langle V\rangle}{\lambda_T^d}\textnormal{Li}_{\frac{d}{2}-1}(\pm e^{\beta\mu})+\frac{\langle N\rangle^2}{\langle V\rangle^2}\Delta^2 V.
\end{equation}

\subsection{Heat capacities}

Heat capacities at constant volume and pressure are

\begin{align}
\label{CV.ideal.quant}
C_V= & \left(\frac{d}{2}+1\right)k_B\beta\langle H\rangle-d\,k_B\beta\,\mu\,\langle N\rangle+k_B\,\beta^2\mu^2\left(\Delta^2N-\frac{\langle N\rangle^2}{\langle V\rangle^2}\Delta^2V\right), \\
\label{CP.ideal.quant}
C_P= & C_V+k_B\beta^2\left(P+\frac{\langle H\rangle}{\langle V\rangle}-\mu\,\frac{\langle N\rangle}{\langle V\rangle}\right)^2\Delta^2V.
\end{align}
Equations \eqref{CV.ideal.quant} and \eqref{CP.ideal.quant} are different from the standard textbook heat capacities where the particle number is fixed, because they also include contibutions due to fluctuations of particle number and volume.

When both volume and particle number are constant, $\frac{\langle N\rangle}{\langle V\rangle}=\frac{\partial P_c}{\partial\mu}$ is constant as well, resulting in an implicit relation between $\beta$ and $\mu$.
The heat capacity under these conditions is

\begin{equation}
C_{V,N}=\left(\frac{dQ}{dT}\right)_{\langle V\rangle,\langle N\rangle}=-k_B\beta^2\left(\frac{d\langle H\rangle}{d\beta}\right)_{\langle V\rangle,\langle N\rangle}=\left(\frac{d}{2}+1\right)k_B\beta\langle H\rangle-\frac{d}{2}\,k_B\beta\langle N\rangle\,\frac{1}{z}\left(\frac{dz}{d\beta}\right)_{\langle V\rangle,\langle N\rangle},
\end{equation}
where $z=e^{\beta\mu}$ is the fugacity.
From the derivative of equation \eqref{av.part.numb.quant} with respect to $\beta$, one obtains the derivative of the fugacity:

\begin{align}
0=\left(\frac{d}{d\beta}\frac{\partial P_c}{\partial\mu}\right)_{\langle V\rangle,\langle N\rangle}= & -\frac{d}{2}\cdot\frac{\textnormal{Li}_{\frac{d}{2}}(z)}{\beta\lambda_T^d}
\pm
\frac{\textnormal{Li}_{\frac{d}{2}-1}(z)}{z\lambda_T^d}\cdot\left(\frac{dz}{d\beta}\right)_{\langle V\rangle,\langle N\rangle} \nonumber \\
= & -\frac{d}{2\beta}\cdot\frac{\langle N\rangle}{\langle V\rangle}+\left(\frac{\Delta^2N}{\langle V\rangle}-\frac{\langle N\rangle^2}{\langle V\rangle^3}\Delta^2V\right)\cdot\frac{1}{z}\left(\frac{dz}{d\beta}\right)_{\langle V\rangle,\langle N\rangle}.
\end{align}
Therefore, the heat capacity at constant volume and particle number is

\begin{equation}
C_{V,N}=\left(\frac{d}{2}+1\right)k_B\beta\langle H\rangle-k_B\,\frac{d^2}{4}\cdot\frac{\langle N\rangle^2\langle V\rangle^2}{\langle V\rangle^2\Delta^2N-\langle N\rangle^2\Delta^2V}.
\end{equation}

When pressure and particle number are both constant, the heat capacity is

\begin{align}
C_{P,N}= & \left(\frac{dQ}{dT}\right)_{P,\langle N\rangle}=-k_B\beta^2\left(\frac{d\langle H\rangle}{d\beta}\right)_{P,\langle N\rangle}-k_B\beta^2P\left(\frac{d\langle V\rangle}{d\beta}\right)_{P,\langle N\rangle} \nonumber \\
= & \left(\frac{d}{2}+1\right)k_B\beta\langle H\rangle-\frac{d}{2}k_B\beta\langle N\rangle\cdot\frac{1}{z}\left(\frac{dz}{d\beta}\right)_{P,\langle N\rangle}-k_B\beta^2\left(\frac{\langle H\rangle}{\langle V\rangle}+P\right)\left(\frac{d\langle V\rangle}{d\beta}\right)_{P,\langle N\rangle} \nonumber \\
= & -\frac{d}{2}k_B\beta^2\left(\frac{dP_c}{d\beta}\right)_{P,\langle N\rangle}\langle V\rangle-k_B\beta^2\left(\frac{d}{2}P_c+P\right)\left(\frac{d\langle V\rangle}{d\beta}\right)_{P,\langle N\rangle}.
\end{align}
The derivative of the average volume is obtained from the following condition

\begin{align}
0=\left(\frac{d\langle N\rangle}{d\beta}\right)_{P,\langle N\rangle}= & \left(\frac{d}{d\beta}\frac{\partial P_c}{\partial\mu}\right)_{P,\langle N\rangle}\langle V\rangle+\frac{\partial P_c}{\partial\mu}\cdot\left(\frac{d\langle V\rangle}{d\beta}\right)_{P,\langle N\rangle} \nonumber \\
= & -\frac{d}{2\beta}\langle N\rangle+\left(\Delta^2N-\frac{\langle N\rangle^2}{\langle V\rangle^2}\Delta^2V\right)\cdot\frac{1}{z}\left(\frac{dz}{d\beta}\right)_{P,\langle N\rangle}+\frac{\langle N\rangle}{\langle V\rangle}\cdot\left(\frac{d\langle V\rangle}{d\beta}\right)_{P,\langle N\rangle},
\end{align}
and the derivative of the fugacity follows from

\begin{equation}
\left(\frac{dP_c}{d\beta}\right)_{P,\langle N\rangle}=-\left(\frac{d}{2}+1\right)\frac{P_c}{\beta}+\frac{\langle N\rangle}{\langle V\rangle}\cdot\frac{1}{\beta z}\left(\frac{dz}{d\beta}\right)_{P,\langle N\rangle}.
\end{equation}
Using the above relations, we obtain

\begin{align}
C_{P,N}= & -\frac{d}{2}k_B\beta\left(\beta\left(\frac{dP_c}{d\beta}\right)_{P,\langle N\rangle}+\frac{d}{2}P_c+P\right)\langle V\rangle \nonumber \\
& +k_B\beta^2\left(\frac{d}{2}P_c+P\right)\left(\beta\left(\frac{dP_c}{d\beta}\right)_{P,\langle N\rangle}+\left(\frac{d}{2}+1\right)P_c\right)\frac{\langle V\rangle^2}{\langle N\rangle^2}\left(\Delta^2N-\frac{\langle N\rangle^2}{\langle V\rangle^2}\Delta^2V\right).
\label{CPN}
\end{align}
Equation \eqref{CPN} is not explicit because of the contribution $\left(\frac{dP_c}{d\beta}\right)_{P,\langle N\rangle}$ which however vanishes when $P=P_c$.

\section{Classical ideal homogeneous gas}

The Legendre transform of the $\mu VT$ for the classical ideal homogeneous gas is the classical limit of the quantum homogeneous gas, namely the limit of small fugacity $e^{\beta\mu}\ll1$, which implies\cite{Wood1992} $\textnormal{Li}_s\big(\pm e^{\beta\mu}\big)\simeq\pm e^{\beta\mu}$. Therefore, thermal quantities computed in the previous section become

\begin{align}
\label{Pc.class.lim}
P_c= & \frac{e^{\beta\mu}}{\beta\lambda_T^d}, \\
\label{N.class.lim}
\langle N\rangle= & \frac{e^{\beta\mu}}{\lambda_T^d}\langle V\rangle, \\
\label{H.class.lim}
\langle H\rangle= & \frac{d}{2\beta}\langle N\rangle, \\
\label{DeltaN.class.lim}
\Delta^2 N= & \langle N\rangle+\frac{\langle N\rangle^2}{\langle V\rangle^2}\Delta^2 V, \\
\label{CV.class.lim}
C_V= & k_B\left(\frac{d}{2}+\left(\frac{d}{2}-\beta\,\mu\right)^2\right)\langle N\rangle, \\
\label{CP.class.lim}
C_P= & C_V+k_B\beta^2\left(P+\left(\frac{d}{2\beta}-\mu\right)\frac{e^{\beta\mu}}{\lambda_T^d}\right)^2\Delta^2V, \\
C_{V,N}= & \frac{d}{2}k_B\langle N\rangle, \\
C_{P,N}= & \frac{d}{2}k_B\left(1-\frac{P}{P_c}\right)\langle N\rangle+\left(\frac{d}{2}+1+\frac{\beta}{P_c}\left(\frac{dP_c}{d\beta}\right)_{P,\langle N\rangle}\right)k_B\frac{P}{P_c}\langle N\rangle. \\
\end{align}
Using the relation $P=P_c$, which implies $\left(\frac{dP_c}{d\beta}\right)_{P,\langle N\rangle}\to0$, we derive the standard heat capacities of the classical ideal gas, i.e. $C_{P,N}=C_{V,N}+k_B\langle N\rangle=\left(\frac{d}{2}+1\right)k_B\langle N\rangle$.

We now focus on the Legendre transform of the $NPT$ ensemble for the classical ideal homogeneous gas. The $NPT$ partition function is \cite{Tuckerman}

\begin{equation} \label{NPT.part.ft.class}
Z_{NPT}=\frac{e^{-\beta\mu_c N}}{V_0\beta P},
\end{equation}
with

\begin{equation} \label{NPT.class.muc}
\mu_c=\frac{1}{\beta}\ln(\beta P\lambda_T^d),
\end{equation}
leading to the $\mu PT$ partition function

\begin{equation} \label{muPT.part.ft.class}
Z_{\mu PT}=\sum_{N_1}^{N_2}e^{\beta\mu N}Z_{NPT}=\frac{e^{\beta N_1(\mu-\mu_c)}-e^{\beta(N_2+1)(\mu-\mu_c)}}{V_0\beta P(1-e^{\beta(\mu-\mu_c)})}.
\end{equation}
Note that the critical chemical potential $\mu_c$ introduced in equation \eqref{NPT.part.ft.class} is different from that in \eqref{ZNPT}, but
their difference is negligible, e.g., if $N\gg1$, so that $Z_{NPT}=e^{-\beta\mu_c N-\ln(V_0\beta P)}\simeq e^{-\beta\mu_c N}$. In particular the two critical chemical potentials agree in the thermodynamic limit, but also for moderately large particle number, like $N\sim 10^3$ as in small thermodynamical systems \cite{Hill,Hill2001,Hill2002}.
Consequently, also thermal quantities computed from the partition functions \eqref{muPT.part.ft.class} and from \eqref{legendre.NPT}, using the critical chemical potential \eqref{NPT.class.muc}, agree within negligible corrections.

Furthermore, the two equations $P=P_c$ and $\mu=\mu_c$ are equivalent. Therefore, as discussied in section \ref{non-comm}, the variances of the volume and of the particle number are both superextensive, in agreement with the linear relations in equations \eqref{part.numb.fluct.hom.gas} and \eqref{DeltaN.class.lim}. The equivalence of the two critical conditions $P=P_c$ and $\mu=\mu_c$ agrees with our general arguments in sections \ref{non-analyticity} and \ref{non-comm}, supporting that these conditions are necessary for the equivalence of the $\mu PT$ ensemble with both the $\mu VT$ and the $NPT$ ones.

\subsection{Thermal averages and fluctuations}

Averages of particle number and of the volume are related by the following equation

\begin{equation} \label{av.vol.class}
\langle V\rangle=\frac{\langle N\rangle+1}{\beta P}=\frac{\lambda_T^d}{e^{\beta\mu_c}}\big(\langle N\rangle+1\big).
\end{equation}
Equation \eqref{av.vol.class} agrees with the prediction of the Legendre transform of the $\mu VT$ ensemble, e.g. equation \eqref{N.class.lim}, in the thermodynamic limit only if $\mu=\mu_c$.

The average energy is

\begin{equation} \label{av.en.class}
\langle H\rangle=\frac{d}{2\beta}\langle N\rangle,
\end{equation}
which agrees with the formula derived from the Legendre transform of the $\mu VT$ ensemble, i.e. equation \eqref{H.class.lim}.

The variances of the particle number and of the volume fulfil the following relation:

\begin{equation}
\Delta^2V=\frac{\langle V\rangle^2}{\big(\langle N\rangle+1\big)^2}\left(\Delta^2N+\langle N\rangle+1\right),
\end{equation}
which agrees with equation \eqref{DeltaN.class.lim} for $\langle N\rangle\gg1$.

\subsection{Heat capacities}

Heat capacities at constant volume and pressure can be computed, respectively, from equations \eqref{CV.legendreNPT} \eqref{CP.legendreNPT} using the critical chemical potential \eqref{NPT.class.muc}:

\begin{align}
\label{CV.class}
C_V= & k_B\,\beta\,\mu\,\langle N\rangle, \\
\label{CP.class}
C_P= & \left(\frac{d}{2}+1\right)k_B\langle N\rangle+\left(\frac{d}{2}+1-\beta\mu\right)^2k_B\,\Delta^2N.
\end{align}
Heat capacities at constant particle number are easily computed:

\begin{align}
C_{V,N}= & \left(\frac{dQ}{dT}\right)_{\langle V\rangle,\langle N\rangle}=-k_B\beta^2\left(\frac{d\langle H\rangle}{d\beta}\right)_{\langle V\rangle,\langle N\rangle}=\frac{d}{2}k_B\langle N\rangle, \\
C_{P,N}= & \left(\frac{dQ}{dT}\right)_{P,\langle N\rangle}=-k_B\beta^2\left(\frac{d\langle H\rangle}{d\beta}+P\,\frac{d\langle V\rangle}{d\beta}\right)_{P,\langle N\rangle}=\left(\frac{d}{2}+1\right)k_B\langle N\rangle+k_B,
\end{align}
which are the standard heat capacities for $\langle N\rangle\gg1$.

\section{Mean-field Bose gas}

This section is devoted to the Bose gas with mean-field interactions, namely with the interaction hamiltonian $\lambda\frac{N}{2V}$. The partition function of the $\mu VT$ ensemble was computed for a class of free Hamiltonians \cite{vandenBerg1984}. Consider here the free Hamiltonian of the ideal homogeneous gas in $d$ dimensions for concreteness. The pressure derived in the $\mu VT$ ensemble is

\begin{equation} \label{press.mf}
P_c^{(\lambda)}(\mu)=\frac{(\mu-\alpha)^2}{2\lambda}+P_c^{(0)}(\alpha),
\end{equation}
where $P_c^{(0)}(\alpha)$ is the critical pressure of the non-interacting gas in equation \eqref{crit.press.hom.gas} with the upper sign and with $\mu$ replaced by $\alpha$, $\alpha$ is zero is $\mu\geqslant\lambda\rho_{\textnormal BEC}$ and is the unique solution of $\alpha+\lambda\partial_\alpha P_c^{(0)}(\beta,\alpha)=\mu$ if $\mu<\lambda\rho_{\textnormal BEC}$, and $\rho_{\textnormal BEC}$ is the critical density of the Bose-Einstein condensation which coincides with that of the non-interacting gas. We shall focus on the regime $\mu<\lambda\rho_{\textnormal BEC}$, that is above the Bose-Einstein condensation temperature.
Using the definition of $\alpha$, the derivatives of the critical pressure \eqref{press.mf} can be expressed as

\begin{align}
\frac{\partial P_c^{(\lambda)}(\mu)}{\partial\beta} & =\frac{\alpha(\mu-\alpha)}{\lambda\beta}-\left(\frac{d}{2}+1\right)\frac{P_c^{(0)}(\alpha)}{\beta}, \\
\frac{\partial P_c^{(\lambda)}(\mu)}{\partial\mu} & =\frac{\mu-\alpha}{\lambda}=\frac{\partial P_c^{(0)}(\alpha)}{\partial\alpha}=\frac{\textnormal{Li}_{\frac{d}{2}}(e^{\beta\alpha})}{\lambda_T^d}, \\
\frac{\partial P_c^{(\lambda)}(\mu)}{\partial\alpha} & =0.
\end{align}

From the definition of $\alpha$, its derivatives satisfy

\begin{align}
\frac{\partial\alpha}{\partial\beta} & =-\lambda\frac{\partial^2P_c^{(0)}(\alpha)}{\partial\beta\partial\alpha}=\frac{\lambda}{\lambda_T^d}\left(\frac{d}{2\beta}\textnormal{Li}_{\frac{d}{2}}(e^{\beta\alpha})-\alpha\textnormal{Li}_{\frac{d}{2}-1}(e^{\beta\alpha})\right)-\lambda\,\frac{\partial\alpha}{\partial\beta}\cdot\frac{\partial^2P_c^{(0)}(\alpha)}{\partial\alpha^2}, \\
\frac{\partial\alpha}{\partial\mu} & =1-\lambda\,\frac{\partial^2P_c^{(0)}(\alpha)}{\partial\mu\partial\alpha}=1-\lambda\,\frac{\partial\alpha}{\partial\mu}\cdot\frac{\partial^2P_c^{(0)}(\alpha)}{\partial\alpha^2}.
\end{align}
Using the expression
\begin{equation}
\frac{\partial^2P_c^{(0)}(\alpha)}{\partial\alpha^2}=\frac{\beta}{\lambda_T^d}\,\textnormal{Li}_{\frac{d}{2}-1}(e^{\beta\alpha}),
\end{equation}
one computes

\begin{align}
\frac{\partial\alpha}{\partial\mu} & =\frac{\lambda_T^d}{\lambda_T^d+\lambda\beta\textnormal{Li}_{\frac{d}{2}-1}(e^{\beta\alpha})}, \\
\frac{\partial\alpha}{\partial\beta} & =\frac{\lambda}{2\beta}\cdot\frac{d\textnormal{Li}_{\frac{d}{2}}(e^{\beta\alpha})-2\beta\alpha\textnormal{Li}_{\frac{d}{2}-1}(e^{\beta\alpha})}{\lambda_T^d+\lambda\beta\textnormal{Li}_{\frac{d}{2}-1}(e^{\beta\alpha})}.
\end{align}

Plugging these derivatives in the general formulas in section \ref{muVT}, we obtain the thermal quantities of the mean-field Bose gas.

\subsection{Thermal averages and fluctuations}

The average particle number and the average energy are, respectively,

\begin{align}
\label{Nav.mf}
\langle N\rangle & =\frac{\mu-\alpha}{\lambda}\langle V\rangle, \\
\langle H\rangle & =\left(\frac{d}{2} \, P_c^{(0)}(\alpha)+\frac{\left(\mu-\alpha\right)^2}{2\lambda}\right)\langle V\rangle=\left(P_c^{(\lambda)}(\mu)+\left(\frac{d}{2}-1\right)P_c^{(0)}(\alpha)\right)\langle V\rangle.
\end{align}

The relation between the variance of the particle number and the variance of the volume is

\begin{equation} \label{varNMF}
\Delta^2N=\left(1-\frac{\partial\alpha}{\partial\mu}\right)\frac{\langle V\rangle}{\lambda\beta}+\left(\frac{\mu-\alpha}{\lambda}\right)^2\Delta^2V.
\end{equation}

\subsection{Heat capacities}

Heat capacities at constant volume and pressure are

\begin{align}
\label{CV.mf}
C_V= & k_B\beta\langle V\rangle\left(\left(\frac{d}{2}+1\right)\frac{d}{2}P_c^{(0)}(\alpha)-\frac{d}{2}\cdot\frac{\alpha\,(\mu-\alpha)}{\lambda}-\left(\frac{\alpha}{\lambda}+\frac{d}{2}\cdot\frac{\mu-\alpha}{\lambda}\right)\beta\,\frac{\partial\alpha}{\partial\beta}\right) \\
\label{CP.mf}
C_P= & C_V+k_B\beta^2\left(P+\frac{\langle H\rangle}{\langle V\rangle}-\frac{\alpha\,(\mu-\alpha)}{\lambda}\right)^2\Delta^2V.
\end{align}

When both volume and particle number are constant, the derivative of equation \eqref{Nav.mf} reads, where $t=e^{\beta\alpha}$,

\begin{equation} \label{NVconst.mf}
0=\left(\frac{d}{d\beta}\frac{\langle N\rangle}{\langle V\rangle}\right)_{\langle V\rangle,\langle N\rangle}=\left(\frac{d}{d\beta}\,\frac{\mu-\alpha}{\lambda}\right)_{\langle V\rangle,\langle N\rangle}=\left(\frac{d}{d\beta}\,\frac{\partial P_c^{(0)}(\alpha)}{\partial\alpha}\right)_{\langle V\rangle,\langle N\rangle}=
\frac{\textnormal{Li}_{\frac{d}{2}-1}(t)}{\lambda_T^dt}\cdot\left(\frac{dt}{d\beta}\right)_{\langle V\rangle,\langle N\rangle}
-\frac{d}{2}\cdot\frac{\textnormal{Li}_{\frac{d}{2}}(t)}{\beta\lambda_T^d}.
\end{equation}
The second equality in \eqref{NVconst.mf} simplifies the heat capacity with both volume and particle number constant:

\begin{equation}
C_{V,N}=\left(\frac{dQ}{dT}\right)_{\langle V\rangle,\langle N\rangle}=-k_B\beta^2\left(\frac{d\langle H\rangle}{d\beta}\right)_{\langle V\rangle,\langle N\rangle}=-k_B\beta^2\langle V\rangle\frac{d}{2}\left(\frac{dP_c^{(0)}(\alpha)}{d\beta}\right)_{\langle V\rangle,\langle N\rangle}.
\end{equation}

Using the following derivative

\begin{equation} \label{derPc0}
\frac{dP_c^{(0)}(\alpha)}{d\beta}=\frac{\textnormal{Li}_{\frac{d}{2}-1}(t)}{\beta\lambda_T^d t}\cdot\frac{dt}{d\beta}
-\left(\frac{d}{2}+1\right)\frac{P_c^{(0)}(\alpha)}{\beta},
\end{equation}
and substituting $\left(\frac{dt}{d\beta}\right)_{\langle V\rangle,\langle N\rangle}$ from equation \eqref{NVconst.mf}, one obtains the heat capacity

\begin{equation}
C_{V,N}=k_B\langle V\rangle\frac{d}{2}\left(
\left(\frac{d}{2}+1\right)\beta P_c^{(0)}(\alpha)
-\frac{d}{2\lambda_T^d}\cdot\frac{\textnormal{Li}^2_{\frac{d}{2}}(t)}{\textnormal{Li}_{\frac{d}{2}-1}(t)}
\right).
\end{equation}

The heat capacity when pressure and particle number are both constant is

\begin{align} \label{CPN.mf}
C_{P,N}= & \left(\frac{dQ}{dT}\right)_{P,\langle N\rangle}=-k_B\beta^2\left(\frac{d\langle H\rangle}{d\beta}\right)_{P,\langle N\rangle}-k_B\beta^2P\left(\frac{d\langle V\rangle}{d\beta}\right)_{P,\langle N\rangle}= \nonumber \\
= & -k_B\beta^2\langle V\rangle\left(
\left(\frac{dP_c^{(\lambda)}(\mu)}{d\beta}\right)_{P,\langle N\rangle}+\left(\frac{d}{2}+1\right)\left(\frac{dP_c^{(0)}(\alpha)}{d\beta}\right)_{P,\langle N\rangle}+\left(\frac{\langle H\rangle}{\langle V\rangle}+P\right)\left(\frac{d\langle V\rangle}{d\beta}\right)_{P,\langle N\rangle}
\right).
\end{align}
First, plug the derivative \eqref{derPc0} into \eqref{CPN.mf}, then eliminate $\left(\frac{d\langle V\rangle}{d\beta}\right)_{P,\langle N\rangle}$ from

\begin{align}
0=\left(\frac{d\langle N\rangle}{d\beta}\right)_{P,\langle N\rangle}=\frac{\textnormal{Li}_{\frac{d}{2}}(t)}{\lambda_T^d}\left(\left(\frac{d\langle V\rangle}{d\beta}\right)_{P,\langle N\rangle}-\frac{d}{2\beta}\langle V\rangle\right)+\frac{\textnormal{Li}_{\frac{d}{2}-1}(t)}{\lambda_T^d t}\cdot\left(\frac{dt}{d\beta}\right)_{P,\langle N\rangle},
\end{align}
and the derivative $\frac{dt}{d\beta}$ from

\begin{align}
\left(\frac{dP_c^{(\lambda)}(\mu)}{d\beta}\right)_{P,\langle N\rangle} & =\frac{\lambda}{2}\left(\frac{d}{d\beta}\left(\frac{\partial P_c^{(0)}(\alpha)}{\partial\alpha}\right)^2\right)_{P,\langle N\rangle}+\left(\frac{dP_c^{(0)}(\alpha)}{d\beta}\right)_{P,\langle N\rangle}= \nonumber \\
& = \left(\frac{dt}{d\beta}\right)_{P,\langle N\rangle}\frac{\textnormal{Li}_{\frac{d}{2}}(t)}{t\lambda_T^d}\left(\frac{1}{\beta}+\frac{\lambda\textnormal{Li}_{\frac{d}{2}-1}(t)}{\lambda_T^d}\right)-\left(\frac{d}{2}+1\right)\frac{P_c^{(0)}(\alpha)}{\beta}-\frac{d\lambda}{2\beta}\left(\frac{\textnormal{Li}_{\frac{d}{2}}(t)}{\lambda_T^d}\right)^2.
\end{align}
Finally, the heat capacity at constant pressure and particle number is

\begin{align}
C_{P,N}= & -k_B\beta\langle V\rangle\left(\beta\left(\frac{dP_c^{(\lambda)}(\mu)}{d\beta}\right)_{P,\langle N\rangle}\frac{\frac{d}{2}\lambda_T^d+\lambda\beta\textnormal{Li}_{\frac{d}{2}-1}(t)}{\lambda_T^d+\lambda\beta\textnormal{Li}_{\frac{d}{2}-1}(t)}+\frac{d}{2}\left(\frac{\langle H\rangle}{\langle V\rangle}+P\right)\frac{\lambda_T^d+2\lambda\beta\textnormal{Li}_{\frac{d}{2}-1}(t)}{\lambda_T^d+\lambda\beta\textnormal{Li}_{\frac{d}{2}-1}(t)}+\right. \nonumber \\
& +\left(\frac{\langle H\rangle}{\langle V\rangle}+P\right)\left(\beta\left(\frac{dP_c^{(\lambda)}(\mu)}{d\beta}\right)_{P,\langle N\rangle}+\left(\frac{d}{2}+1\right)P_c^{(\lambda)}(\mu)\right)\left(\frac{\lambda_T^d}{\textnormal{Li}_{\frac{d}{2}}(t)}\right)^2\frac{\textnormal{Li}_{\frac{d}{2}-1}(t)}{\lambda_T^d+\lambda\beta\textnormal{Li}_{\frac{d}{2}-1}(t)}+ \nonumber \\
& \left.-\left(\frac{d^2}{4}-1\right)P_c^{(\lambda)}(\mu)\frac{\lambda\beta\textnormal{Li}_{\frac{d}{2}-1}(t)}{\lambda_T^d+\lambda\beta\textnormal{Li}_{\frac{d}{2}-1}(t)}+\frac{d}{2}\left(\frac{d}{2}-1\right)\left(\frac{\lambda_T^d}{\lambda\textnormal{Li}_{\frac{d}{2}}(t)}\right)^2\frac{\lambda\lambda_T^d}{\lambda_T^d+\lambda\beta\textnormal{Li}_{\frac{d}{2}-1}(t)}\right).
\end{align}

All the expressions for the mean field Bose gas recover those for the ideal homogeneous gas, when $\lambda\to0$, recalling that $\alpha\to\mu$ and $t\to z$ in this limit.

\section{Conclusions}

The $\mu PT$ statistical ensemble, analysed here, describes equilibrium systems which exchange energy, particles, and volume with the surrounding.
This ensemble finds applications in the thermodynamics of small systems, like nanothermodynamics, systems with long-range interactions, and systems confined within a porous and elastic membranes.
The statistics of the volume and the particle number do not depend on the specific model, i.e. on the Hamiltonian, as a consequence of the Legendre transforms performed on all the extensive quantities. Another peculiarity of the $\mu PT$ ensemble is that values of pressure and chemical potential agree with $\mu VT$ and $NPT$ ensembles only around non-analyticity points $P=P_c$ and $\mu=\mu_c$ at which, however, fluctuations of volume and particle number are superextensive. The constraint on intensive parameters, in agreement with the thermodynamical Gibbs-Duhem equation, also emerges from the requirement that the order of the Legendre transforms with respect to the volume and to the particle number do not alter thermodynamic relations in the $\mu PT$ ensemble. Therefore, the breakdown of the Gibbs-Duhem equation is related to the non-commutativity of Legendre transforms for large system size. The order of the Legendre transforms could be indicated by the specific model or by experimental conditions. Quantum and classical ideal gases, and a quantum mean-field Bose gas are discussed in order to exemplify the general featurs of the $\mu PT$ ensemble.


\section*{Acknowledgements}
U.~M. acknowledges interesting and helpful discussions with Fabio Staniscia and Andrea Trombettoni, and
is financially supported by the European Union's Horizon 2020 research and innovation programme under the Marie Sk\l odowska-Curie grant agreement No. 754496 - FELLINI.

\appendix

\section{Comparison with other ensembles} \label{app}

Some fundamental differences of the $\mu PT$ ensemble with respect to other statistical ensembles stem from the fact that all possible Legendre transforms with respect to internal quantities have been computed. The $\mu PT$ partition function thus depends only on intensive parameters unless other internal quantities $\{X_l\}_l$, even though fixed, contribute to energy, particles, and volume exchanges.

One peculiarity of the $\mu PT$ ensemble is that the above analysis of the volume statistics and of the particle number statistics is general, and does not rely upon the specific model. The reason for such generality is the dependence of the $\mu VT$ thermodynamic potential and of the $NPT$ thermodynamic potential on a single extensive parameter, i.e. $V$ and $N$ respectively. The same general behaviour does not hold for other statistical ensemble where the thermodynamic potential depends on more that one extensive quantity.

For instance, both the $\mu VT$ ensemble and the $NPT$ ensemble are derived from Legendre transforms of the $NVT$ ensemble which, together with its thermodynamic potential $F=-pV+\mu N$, depends on two extensive, fixed quantities, e.g. $N$ and $V$. Thus, the intensive quantities $P$ and $\mu$ can depend on the ratio $N/V$, resulting in a non-linear dependence on $N$ and yet an extensive Helmholtz free energy. An example is the ideal homogeneous classical gas in $d$ dimensions, whose $NVT$ partition function is \cite{Tuckerman}

\begin{equation}
Z_{NVT}=\frac{V^N}{N!\lambda_T^{dN}},
\end{equation}
where $\lambda_T=\sqrt{2\pi h^2\beta/m}$ is the thermal wavelength, and with the Helmholtz free energy

\begin{equation}
F=-\frac{\ln Z_{NVT}}{\beta}=-\frac{N}{\beta}\ln\frac{eV}{N\lambda_T^d}+\mathcal{O}(\ln N)=-PV+\mu N+\mathcal{O}(\ln N),
\end{equation}
where the Stirling's approximation $\ln N!=N\ln(N/e)+\mathcal{O}(\ln N)$ has been used. The pressure and the chemical potential are derivatives of the Helmholtz free energy:

\begin{equation} \label{parNVT}
P=-\frac{\partial F}{\partial V}=\frac{N}{\beta V}, \qquad
\mu=\frac{\partial F}{\partial N}=\frac{1}{\beta}\ln\frac{N\lambda_T^d}{V}.
\end{equation}

The possible non-linear dependence of the $NVT$ thermodynamic parameter on $N$ and on $V$, although $F$ is extensive, results in model-dependent particle number statistics of the $\mu VT$ ensemble and volume statistics of the $NPT$ ensemble, after the respective Legendre transforms. This is not the case for volume statistics and particle number statistics in the $\mu PT$ ensemble, because they are derived from Legendre transforms of the $\mu VT$ ensemble and of the $NPT$ ensemble respectively, which depend only on a single extensive, fixed parameter, namely $V$ and $N$ respectively. Thus, the $\mu VT$ and the $NPT$ thermodynamic potentials, in order to be extensive, can only have a linear dependence on $V$ and $N$ respectively. This implies simple and general volume and particle number statistics in the $\mu PT$ ensemble.

Another difference between the $\mu PT$ ensemble and others deals with the structure of Legendre transforms. When Legendre transforms are applied to derive statistical ensembles from others, e.g. $NVT$ from $NVE$, $\mu VT$ from $NVT$, or $NPT$ from $NVT$, there are intensive quantities of the original ensemble that are not control parameters, but can be derived from derivatives of thermodynamic potentials (see the last column of table \ref{thermodyn.pot} and equation \eqref{parNVT}). These intensive parameters depend on the control parameters that define the ensemble (as the ensemble names denote): among these control parameters there are also extensive quantities statistically conjugated to intensive parameters that are not control parameters. After a Legendre transform, an extensive, previously fixed, quantity becomes a stochastic variable and the dependence of its expectation value from the control parameters is the inverse function of its intensive statistically conjugated variable before the Legendre transform.

The aforementioned construction is not straightforward in the Legendre transform of the $\mu VT$ ensemble with respect to the volume and in the Legendre transform of the $NPT$ ensemble with respect to the particle number, both leading to the $\mu PT$ ensemble. The reason is that pressure in the $\mu VT$ ensemble and chemical potential in the $NPT$ ensemble only depend on other intensive parameters and not on the volume or on the particle number respectively. This is the only way the thermodynamic potentials of the $\mu VT$ and $NPT$ ensembles, i.e. $PV$ and $-\mu N$ respectively, can be extensive. Nevertheless, the pressure and the chemical potential in the $\mu PT$ ensemble are independent parameters. It is therefore natural to expect the emergence of a relation constraining the intensive parameters of the $\mu PT$ ensemble from ensemble equivalence, as discussed in section \ref{non-comm}.



\begin{thebibliography}{55}%
\makeatletter
\providecommand \@ifxundefined [1]{%
 \@ifx{#1\undefined}
}%
\providecommand \@ifnum [1]{%
 \ifnum #1\expandafter \@firstoftwo
 \else \expandafter \@secondoftwo
 \fi
}%
\providecommand \@ifx [1]{%
 \ifx #1\expandafter \@firstoftwo
 \else \expandafter \@secondoftwo
 \fi
}%
\providecommand \natexlab [1]{#1}%
\providecommand \enquote  [1]{``#1''}%
\providecommand \bibnamefont  [1]{#1}%
\providecommand \bibfnamefont [1]{#1}%
\providecommand \citenamefont [1]{#1}%
\providecommand \href@noop [0]{\@secondoftwo}%
\providecommand \href [0]{\begingroup \@sanitize@url \@href}%
\providecommand \@href[1]{\@@startlink{#1}\@@href}%
\providecommand \@@href[1]{\endgroup#1\@@endlink}%
\providecommand \@sanitize@url [0]{\catcode `\\12\catcode `\$12\catcode
  `\&12\catcode `\#12\catcode `\^12\catcode `\_12\catcode `\%12\relax}%
\providecommand \@@startlink[1]{}%
\providecommand \@@endlink[0]{}%
\providecommand \url  [0]{\begingroup\@sanitize@url \@url }%
\providecommand \@url [1]{\endgroup\@href {#1}{\urlprefix }}%
\providecommand \urlprefix  [0]{URL }%
\providecommand \Eprint [0]{\href }%
\providecommand \doibase [0]{https://doi.org/}%
\providecommand \selectlanguage [0]{\@gobble}%
\providecommand \bibinfo  [0]{\@secondoftwo}%
\providecommand \bibfield  [0]{\@secondoftwo}%
\providecommand \translation [1]{[#1]}%
\providecommand \BibitemOpen [0]{}%
\providecommand \bibitemStop [0]{}%
\providecommand \bibitemNoStop [0]{.\EOS\space}%
\providecommand \EOS [0]{\spacefactor3000\relax}%
\providecommand \BibitemShut  [1]{\csname bibitem#1\endcsname}%
\let\auto@bib@innerbib\@empty
\bibitem [{\citenamefont {Landau}\ and\ \citenamefont
  {Binder}(2000)}]{LandauBinder}%
  \BibitemOpen
  \bibfield  {author} {\bibinfo {author} {\bibfnamefont {D.~P.}\ \bibnamefont
  {Landau}}\ and\ \bibinfo {author} {\bibfnamefont {K.}~\bibnamefont
  {Binder}},\ }\href@noop {} {\emph {\bibinfo {title} {{A Guide to Monte Carlo
  Simulations in Statistical Physics}}}}\ (\bibinfo  {publisher} {Cambridge
  University Press},\ \bibinfo {year} {2000})\BibitemShut {NoStop}%
\bibitem [{\citenamefont {Galiba}\ \emph {et~al.}(2017)\citenamefont {Galiba},
  \citenamefont {Duignan}, \citenamefont {Mistelib}, \citenamefont {Baer},
  \citenamefont {Schenter}, \citenamefont {Hutterb},\ and\ \citenamefont
  {Mundy}}]{Galib2017}%
  \BibitemOpen
  \bibfield  {author} {\bibinfo {author} {\bibfnamefont {M.}~\bibnamefont
  {Galiba}}, \bibinfo {author} {\bibfnamefont {T.~T.}\ \bibnamefont {Duignan}},
  \bibinfo {author} {\bibfnamefont {Y.}~\bibnamefont {Mistelib}}, \bibinfo
  {author} {\bibfnamefont {M.~D.}\ \bibnamefont {Baer}}, \bibinfo {author}
  {\bibfnamefont {G.~K.}\ \bibnamefont {Schenter}}, \bibinfo {author}
  {\bibfnamefont {J.}~\bibnamefont {Hutterb}},\ and\ \bibinfo {author}
  {\bibfnamefont {C.~J.}\ \bibnamefont {Mundy}},\ }\href@noop {} {\bibfield
  {journal} {\bibinfo  {journal} {J. Chem. Phys.}\ }\textbf {\bibinfo {volume}
  {146}},\ \bibinfo {pages} {244501} (\bibinfo {year} {2017})}\BibitemShut
  {NoStop}%
\bibitem [{\citenamefont {Gallavotti}(1999)}]{Gallavotti}%
  \BibitemOpen
  \bibfield  {author} {\bibinfo {author} {\bibfnamefont {G.}~\bibnamefont
  {Gallavotti}},\ }\href@noop {} {\emph {\bibinfo {title} {{Statistical
  Mechanics - A Short Treatise}}}}\ (\bibinfo  {publisher} {Springer},\
  \bibinfo {year} {1999})\BibitemShut {NoStop}%
\bibitem [{\citenamefont {Attard}(2002)}]{Attard}%
  \BibitemOpen
  \bibfield  {author} {\bibinfo {author} {\bibfnamefont {P.}~\bibnamefont
  {Attard}},\ }\href@noop {} {\emph {\bibinfo {title} {{Thermodynamics and
  statistical mechanics-Equilibrium by entropy maximisation}}}}\ (\bibinfo
  {publisher} {Academic Press},\ \bibinfo {year} {2002})\BibitemShut {NoStop}%
\bibitem [{\citenamefont {Tuckerman}(2010)}]{Tuckerman}%
  \BibitemOpen
  \bibfield  {author} {\bibinfo {author} {\bibfnamefont {M.~E.}\ \bibnamefont
  {Tuckerman}},\ }\href@noop {} {\emph {\bibinfo {title} {{Statistical
  Mechanics: Theory and Molecular Simulation}}}}\ (\bibinfo  {publisher}
  {Oxford University Press},\ \bibinfo {year} {2010})\BibitemShut {NoStop}%
\bibitem [{\citenamefont {Zia}\ \emph {et~al.}(2009)\citenamefont {Zia},
  \citenamefont {Redish},\ and\ \citenamefont {McKay}}]{Zia2009}%
  \BibitemOpen
  \bibfield  {author} {\bibinfo {author} {\bibfnamefont {R.~K.~P.}\
  \bibnamefont {Zia}}, \bibinfo {author} {\bibfnamefont {E.~F.}\ \bibnamefont
  {Redish}},\ and\ \bibinfo {author} {\bibfnamefont {S.~R.}\ \bibnamefont
  {McKay}},\ }\href@noop {} {\bibfield  {journal} {\bibinfo  {journal} {Am. J.
  Phys.}\ }\textbf {\bibinfo {volume} {77}},\ \bibinfo {pages} {614} (\bibinfo
  {year} {2009})}\BibitemShut {NoStop}%
\bibitem [{\citenamefont {Jaynes}(1957{\natexlab{a}})}]{Jaynes1957-1}%
  \BibitemOpen
  \bibfield  {author} {\bibinfo {author} {\bibfnamefont {E.~T.}\ \bibnamefont
  {Jaynes}},\ }\href@noop {} {\bibfield  {journal} {\bibinfo  {journal} {Phys.
  Rev.}\ }\textbf {\bibinfo {volume} {106}},\ \bibinfo {pages} {620} (\bibinfo
  {year} {1957}{\natexlab{a}})}\BibitemShut {NoStop}%
\bibitem [{\citenamefont {Jaynes}(1957{\natexlab{b}})}]{Jaynes1957-2}%
  \BibitemOpen
  \bibfield  {author} {\bibinfo {author} {\bibfnamefont {E.~T.}\ \bibnamefont
  {Jaynes}},\ }\href@noop {} {\bibfield  {journal} {\bibinfo  {journal} {Phys.
  Rev.}\ }\textbf {\bibinfo {volume} {108}},\ \bibinfo {pages} {171} (\bibinfo
  {year} {1957}{\natexlab{b}})}\BibitemShut {NoStop}%
\bibitem [{\citenamefont {Hill}(1994)}]{Hill}%
  \BibitemOpen
  \bibfield  {author} {\bibinfo {author} {\bibfnamefont {T.~L.}\ \bibnamefont
  {Hill}},\ }\href@noop {} {\emph {\bibinfo {title} {{Thermodynamics of small
  systems}}}}\ (\bibinfo  {publisher} {Dove Publications, INC. New York},\
  \bibinfo {year} {1994})\BibitemShut {NoStop}%
\bibitem [{\citenamefont {Guggenheim}(1967)}]{Guggenheim}%
  \BibitemOpen
  \bibfield  {author} {\bibinfo {author} {\bibfnamefont {E.~A.}\ \bibnamefont
  {Guggenheim}},\ }\href@noop {} {\emph {\bibinfo {title} {{Thermodynamics An
  Advanced Treatment for Chemists and Physicists}}}}\ (\bibinfo  {publisher}
  {Elsevier Science Publishers},\ \bibinfo {year} {1967})\BibitemShut {NoStop}%
\bibitem [{\citenamefont {Campa}\ \emph {et~al.}(2018)\citenamefont {Campa},
  \citenamefont {Casetti}, \citenamefont {Latella}, \citenamefont
  {P\'erez-Madrid},\ and\ \citenamefont {Ruffo}}]{Campa2018}%
  \BibitemOpen
  \bibfield  {author} {\bibinfo {author} {\bibfnamefont {A.}~\bibnamefont
  {Campa}}, \bibinfo {author} {\bibfnamefont {L.}~\bibnamefont {Casetti}},
  \bibinfo {author} {\bibfnamefont {I.}~\bibnamefont {Latella}}, \bibinfo
  {author} {\bibfnamefont {A.}~\bibnamefont {P\'erez-Madrid}},\ and\ \bibinfo
  {author} {\bibfnamefont {S.}~\bibnamefont {Ruffo}},\ }\href@noop {}
  {\bibfield  {journal} {\bibinfo  {journal} {Entropy}\ }\textbf {\bibinfo
  {volume} {20}},\ \bibinfo {pages} {12} (\bibinfo {year} {2018})}\BibitemShut
  {NoStop}%
\bibitem [{\citenamefont {Hill}(2002)}]{Hill2002}%
  \BibitemOpen
  \bibfield  {author} {\bibinfo {author} {\bibfnamefont {T.~L.}\ \bibnamefont
  {Hill}},\ }\href@noop {} {\bibfield  {journal} {\bibinfo  {journal} {Nano
  Lett.}\ }\textbf {\bibinfo {volume} {2}},\ \bibinfo {pages} {609} (\bibinfo
  {year} {2002})}\BibitemShut {NoStop}%
\bibitem [{\citenamefont {Calabrese}\ \emph {et~al.}(2019)\citenamefont
  {Calabrese}, \citenamefont {Rondoni},\ and\ \citenamefont
  {Porporato}}]{Calabrese2019}%
  \BibitemOpen
  \bibfield  {author} {\bibinfo {author} {\bibfnamefont {S.}~\bibnamefont
  {Calabrese}}, \bibinfo {author} {\bibfnamefont {L.}~\bibnamefont {Rondoni}},\
  and\ \bibinfo {author} {\bibfnamefont {A.}~\bibnamefont {Porporato}}}
  (\bibinfo {year} {2019}),\ \bibinfo {note} {preprint
  arXiv:1909.09479}\BibitemShut {NoStop}%
\bibitem [{\citenamefont {Hill}(2001)}]{Hill2001}%
  \BibitemOpen
  \bibfield  {author} {\bibinfo {author} {\bibfnamefont {T.~L.}\ \bibnamefont
  {Hill}},\ }\href@noop {} {\bibfield  {journal} {\bibinfo  {journal} {Nano
  Lett.}\ }\textbf {\bibinfo {volume} {1}},\ \bibinfo {pages} {273} (\bibinfo
  {year} {2001})}\BibitemShut {NoStop}%
\bibitem [{\citenamefont {Chamberlin}(2000)}]{Chamberlin2000}%
  \BibitemOpen
  \bibfield  {author} {\bibinfo {author} {\bibfnamefont {R.~V.}\ \bibnamefont
  {Chamberlin}},\ }\href@noop {} {\bibfield  {journal} {\bibinfo  {journal}
  {Nature}\ }\textbf {\bibinfo {volume} {408}},\ \bibinfo {pages} {6810}
  (\bibinfo {year} {2000})}\BibitemShut {NoStop}%
\bibitem [{\citenamefont {Qian}(2012)}]{Qian2012}%
  \BibitemOpen
  \bibfield  {author} {\bibinfo {author} {\bibfnamefont {H.}~\bibnamefont
  {Qian}},\ }\href@noop {} {\bibfield  {journal} {\bibinfo  {journal} {J. Biol.
  Phys.}\ }\textbf {\bibinfo {volume} {38}},\ \bibinfo {pages} {201} (\bibinfo
  {year} {2012})}\BibitemShut {NoStop}%
\bibitem [{\citenamefont {Chamberlin}(2015)}]{Chamberlin2015}%
  \BibitemOpen
  \bibfield  {author} {\bibinfo {author} {\bibfnamefont {R.~V.}\ \bibnamefont
  {Chamberlin}},\ }\href@noop {} {\bibfield  {journal} {\bibinfo  {journal}
  {Entropy}\ }\textbf {\bibinfo {volume} {17}},\ \bibinfo {pages} {52}
  (\bibinfo {year} {2015})}\BibitemShut {NoStop}%
\bibitem [{\citenamefont {Bedeaux}\ and\ \citenamefont
  {Kjelstrup}(2018)}]{Bedeaux2018}%
  \BibitemOpen
  \bibfield  {author} {\bibinfo {author} {\bibfnamefont {D.}~\bibnamefont
  {Bedeaux}}\ and\ \bibinfo {author} {\bibfnamefont {S.}~\bibnamefont
  {Kjelstrup}},\ }\href@noop {} {\bibfield  {journal} {\bibinfo  {journal}
  {Nano Lett.}\ }\textbf {\bibinfo {volume} {707}},\ \bibinfo {pages} {40}
  (\bibinfo {year} {2018})}\BibitemShut {NoStop}%
\bibitem [{\citenamefont {Latella}\ \emph {et~al.}(2017)\citenamefont
  {Latella}, \citenamefont {P\'erez-Madrid}, \citenamefont {Campa},
  \citenamefont {Casetti},\ and\ \citenamefont {Ruffo}}]{Latella2017}%
  \BibitemOpen
  \bibfield  {author} {\bibinfo {author} {\bibfnamefont {I.}~\bibnamefont
  {Latella}}, \bibinfo {author} {\bibfnamefont {A.}~\bibnamefont
  {P\'erez-Madrid}}, \bibinfo {author} {\bibfnamefont {A.}~\bibnamefont
  {Campa}}, \bibinfo {author} {\bibfnamefont {L.}~\bibnamefont {Casetti}},\
  and\ \bibinfo {author} {\bibfnamefont {S.}~\bibnamefont {Ruffo}},\
  }\href@noop {} {\bibfield  {journal} {\bibinfo  {journal} {Phys. Rev. E}\
  }\textbf {\bibinfo {volume} {95}},\ \bibinfo {pages} {012140} (\bibinfo
  {year} {2017})}\BibitemShut {NoStop}%
\bibitem [{\citenamefont {Campa}\ \emph {et~al.}(2020)\citenamefont {Campa},
  \citenamefont {Casetti}, \citenamefont {Latella},\ and\ \citenamefont
  {Ruffo}}]{Campa2020}%
  \BibitemOpen
  \bibfield  {author} {\bibinfo {author} {\bibfnamefont {A.}~\bibnamefont
  {Campa}}, \bibinfo {author} {\bibfnamefont {L.}~\bibnamefont {Casetti}},
  \bibinfo {author} {\bibfnamefont {I.}~\bibnamefont {Latella}},\ and\ \bibinfo
  {author} {\bibfnamefont {S.}~\bibnamefont {Ruffo}},\ }\href@noop {}
  {\bibfield  {journal} {\bibinfo  {journal} {J. Stat. Mech.}\ ,\ \bibinfo
  {pages} {014004}} (\bibinfo {year} {2020})}\BibitemShut {NoStop}%
\bibitem [{\citenamefont {Abramowitz}\ and\ \citenamefont
  {Stegun}(1970)}]{AbramowitzStegun}%
  \BibitemOpen
  \bibinfo {editor} {\bibfnamefont {M.}~\bibnamefont {Abramowitz}}\ and\
  \bibinfo {editor} {\bibfnamefont {I.~A.}\ \bibnamefont {Stegun}},\ eds.,\
  \href@noop {} {\emph {\bibinfo {title} {{Handbook of Mathematical Functions
  With Formulas, Graphs, and Mathematical Tables}}}}\ (\bibinfo  {publisher}
  {Dover, New York},\ \bibinfo {year} {1970})\BibitemShut {NoStop}%
\bibitem [{\citenamefont {Benden}\ and\ \citenamefont
  {Orszag}(1999)}]{BenderOrszag}%
  \BibitemOpen
  \bibfield  {author} {\bibinfo {author} {\bibfnamefont {C.~M.}\ \bibnamefont
  {Benden}}\ and\ \bibinfo {author} {\bibfnamefont {S.~A.}\ \bibnamefont
  {Orszag}},\ }\href@noop {} {\emph {\bibinfo {title} {{Advanced mathematical
  methods for scientists and engineers}}}}\ (\bibinfo  {publisher} {Springer,
  New York},\ \bibinfo {year} {1999})\BibitemShut {NoStop}%
\bibitem [{\citenamefont {Bleistein}\ and\ \citenamefont
  {Handelsman}(2010)}]{BleisteinHandelsman}%
  \BibitemOpen
  \bibfield  {author} {\bibinfo {author} {\bibfnamefont {N.}~\bibnamefont
  {Bleistein}}\ and\ \bibinfo {author} {\bibfnamefont {R.~A.}\ \bibnamefont
  {Handelsman}},\ }\href@noop {} {\emph {\bibinfo {title} {{Asymptotic
  Expansions of Integrals}}}}\ (\bibinfo  {publisher} {Dover, New York},\
  \bibinfo {year} {2010})\BibitemShut {NoStop}%
\bibitem [{\citenamefont {Gilmore}(1985)}]{Gilmore1985}%
  \BibitemOpen
  \bibfield  {author} {\bibinfo {author} {\bibfnamefont {R.}~\bibnamefont
  {Gilmore}},\ }\href@noop {} {\bibfield  {journal} {\bibinfo  {journal} {Phys.
  Rev. A}\ }\textbf {\bibinfo {volume} {31}},\ \bibinfo {pages} {3237}
  (\bibinfo {year} {1985})}\BibitemShut {NoStop}%
\bibitem [{\citenamefont {Falcioni}\ \emph {et~al.}(2011)\citenamefont
  {Falcioni}, \citenamefont {Villamaina}, \citenamefont {Vulpiani},
  \citenamefont {Puglisi}, \citenamefont {A.},\ and\ \citenamefont
  {Sarracino}}]{Falcioni2011}%
  \BibitemOpen
  \bibfield  {author} {\bibinfo {author} {\bibfnamefont {M.}~\bibnamefont
  {Falcioni}}, \bibinfo {author} {\bibfnamefont {D.}~\bibnamefont
  {Villamaina}}, \bibinfo {author} {\bibfnamefont {A.}~\bibnamefont
  {Vulpiani}}, \bibinfo {author} {\bibnamefont {Puglisi}}, \bibinfo {author}
  {\bibnamefont {A.}},\ and\ \bibinfo {author} {\bibfnamefont {A.}~\bibnamefont
  {Sarracino}},\ }\href@noop {} {\bibfield  {journal} {\bibinfo  {journal} {Am.
  J. Phys.}\ }\textbf {\bibinfo {volume} {79}},\ \bibinfo {pages} {7777}
  (\bibinfo {year} {2011})}\BibitemShut {NoStop}%
\bibitem [{\citenamefont {Davis}\ and\ \citenamefont
  {Guti\'errez}(2012)}]{Davis2012}%
  \BibitemOpen
  \bibfield  {author} {\bibinfo {author} {\bibfnamefont {S.}~\bibnamefont
  {Davis}}\ and\ \bibinfo {author} {\bibfnamefont {G.}~\bibnamefont
  {Guti\'errez}},\ }\href@noop {} {\bibfield  {journal} {\bibinfo  {journal}
  {Phys. Rev. E}\ }\textbf {\bibinfo {volume} {86}},\ \bibinfo {pages} {051136}
  (\bibinfo {year} {2012})}\BibitemShut {NoStop}%
\bibitem [{\citenamefont {Hiura}\ and\ \citenamefont {S.}(2018)}]{Hiura2018}%
  \BibitemOpen
  \bibfield  {author} {\bibinfo {author} {\bibfnamefont {K.}~\bibnamefont
  {Hiura}}\ and\ \bibinfo {author} {\bibfnamefont {S.}~\bibnamefont {S.}},\
  }\href@noop {} {\bibfield  {journal} {\bibinfo  {journal} {J. Stat. Phys.}\
  }\textbf {\bibinfo {volume} {173}},\ \bibinfo {pages} {285} (\bibinfo {year}
  {2018})}\BibitemShut {NoStop}%
\bibitem [{\citenamefont {Cramer}(1946)}]{Cramer}%
  \BibitemOpen
  \bibfield  {author} {\bibinfo {author} {\bibfnamefont {H.}~\bibnamefont
  {Cramer}},\ }\href@noop {} {\emph {\bibinfo {title} {{Mathematical methods of
  statistics}}}}\ (\bibinfo  {publisher} {Princeton University Press},\
  \bibinfo {year} {1946})\BibitemShut {NoStop}%
\bibitem [{\citenamefont {Helstrom}(1976)}]{Helstrom}%
  \BibitemOpen
  \bibfield  {author} {\bibinfo {author} {\bibfnamefont {C.~W.}\ \bibnamefont
  {Helstrom}},\ }\href@noop {} {\emph {\bibinfo {title} {{Quantum Detection and
  Estimation Theory}}}}\ (\bibinfo  {publisher} {Academic Press, New York},\
  \bibinfo {year} {1976})\BibitemShut {NoStop}%
\bibitem [{\citenamefont {Holevo}(2001)}]{Holevo}%
  \BibitemOpen
  \bibfield  {author} {\bibinfo {author} {\bibfnamefont {A.~S.}\ \bibnamefont
  {Holevo}},\ }\href@noop {} {\emph {\bibinfo {title} {{Statistical Structure
  of Quantum Theory}}}},\ \bibinfo {series} {Lect. Not. Phys.}, Vol.~\bibinfo
  {volume} {61}\ (\bibinfo  {publisher} {Spring, Berling},\ \bibinfo {year}
  {2001})\BibitemShut {NoStop}%
\bibitem [{\citenamefont {Weinhold}(1975)}]{Weinhold1975}%
  \BibitemOpen
  \bibfield  {author} {\bibinfo {author} {\bibfnamefont {F.}~\bibnamefont
  {Weinhold}},\ }\href@noop {} {\bibfield  {journal} {\bibinfo  {journal} {J.
  Chem. Phys.}\ }\textbf {\bibinfo {volume} {63}},\ \bibinfo {pages} {2488}
  (\bibinfo {year} {1975})}\BibitemShut {NoStop}%
\bibitem [{\citenamefont {Salamon}\ \emph {et~al.}(1984)\citenamefont
  {Salamon}, \citenamefont {Nulton},\ and\ \citenamefont
  {Ihrig}}]{Salamon1984}%
  \BibitemOpen
  \bibfield  {author} {\bibinfo {author} {\bibfnamefont {P.}~\bibnamefont
  {Salamon}}, \bibinfo {author} {\bibfnamefont {J.}~\bibnamefont {Nulton}},\
  and\ \bibinfo {author} {\bibfnamefont {E.}~\bibnamefont {Ihrig}},\
  }\href@noop {} {\bibfield  {journal} {\bibinfo  {journal} {J. Chem. Phys.}\
  }\textbf {\bibinfo {volume} {80}},\ \bibinfo {pages} {436} (\bibinfo {year}
  {1984})}\BibitemShut {NoStop}%
\bibitem [{\citenamefont {Di\'osi}\ \emph {et~al.}(1984)\citenamefont
  {Di\'osi}, \citenamefont {Forg\'acs}, \citenamefont {Luk\'acs},\ and\
  \citenamefont {Frisch}}]{Diosi1984}%
  \BibitemOpen
  \bibfield  {author} {\bibinfo {author} {\bibfnamefont {L.}~\bibnamefont
  {Di\'osi}}, \bibinfo {author} {\bibfnamefont {G.}~\bibnamefont {Forg\'acs}},
  \bibinfo {author} {\bibfnamefont {B.}~\bibnamefont {Luk\'acs}},\ and\
  \bibinfo {author} {\bibfnamefont {H.~L.}\ \bibnamefont {Frisch}},\
  }\href@noop {} {\bibfield  {journal} {\bibinfo  {journal} {Phys. Rev. A}\
  }\textbf {\bibinfo {volume} {29}},\ \bibinfo {pages} {3343} (\bibinfo {year}
  {1984})}\BibitemShut {NoStop}%
\bibitem [{\citenamefont {Nulton}\ and\ \citenamefont
  {Salamon}(1985)}]{Nulton1985}%
  \BibitemOpen
  \bibfield  {author} {\bibinfo {author} {\bibfnamefont {J.~D.}\ \bibnamefont
  {Nulton}}\ and\ \bibinfo {author} {\bibfnamefont {P.}~\bibnamefont
  {Salamon}},\ }\href@noop {} {\bibfield  {journal} {\bibinfo  {journal} {Phys.
  Rev. A}\ }\textbf {\bibinfo {volume} {31}},\ \bibinfo {pages} {2520}
  (\bibinfo {year} {1985})}\BibitemShut {NoStop}%
\bibitem [{\citenamefont {Janyszek}(1986{\natexlab{a}})}]{Janyszek1986}%
  \BibitemOpen
  \bibfield  {author} {\bibinfo {author} {\bibfnamefont {H.}~\bibnamefont
  {Janyszek}},\ }\href@noop {} {\bibfield  {journal} {\bibinfo  {journal} {Rep.
  Math. Phys.}\ }\textbf {\bibinfo {volume} {24}},\ \bibinfo {pages} {1}
  (\bibinfo {year} {1986}{\natexlab{a}})}\BibitemShut {NoStop}%
\bibitem [{\citenamefont {Janyszek}(1986{\natexlab{b}})}]{Janyszek1986-2}%
  \BibitemOpen
  \bibfield  {author} {\bibinfo {author} {\bibfnamefont {H.}~\bibnamefont
  {Janyszek}},\ }\href@noop {} {\bibfield  {journal} {\bibinfo  {journal} {Rep.
  Math. Phys.}\ }\textbf {\bibinfo {volume} {24}},\ \bibinfo {pages} {11}
  (\bibinfo {year} {1986}{\natexlab{b}})}\BibitemShut {NoStop}%
\bibitem [{\citenamefont {Janyszek}\ and\ \citenamefont
  {Mruga\l{}a}(1989)}]{Janyszek1989-2}%
  \BibitemOpen
  \bibfield  {author} {\bibinfo {author} {\bibfnamefont {H.}~\bibnamefont
  {Janyszek}}\ and\ \bibinfo {author} {\bibfnamefont {R.}~\bibnamefont
  {Mruga\l{}a}},\ }\href@noop {} {\bibfield  {journal} {\bibinfo  {journal}
  {Rep. Math. Phys.}\ }\textbf {\bibinfo {volume} {27}},\ \bibinfo {pages}
  {145} (\bibinfo {year} {1989})}\BibitemShut {NoStop}%
\bibitem [{\citenamefont {Ruppeiner}(1995)}]{Ruppeiner1995}%
  \BibitemOpen
  \bibfield  {author} {\bibinfo {author} {\bibfnamefont {G.}~\bibnamefont
  {Ruppeiner}},\ }\href@noop {} {\bibfield  {journal} {\bibinfo  {journal}
  {Rev. Mod. Phys.}\ }\textbf {\bibinfo {volume} {67}},\ \bibinfo {pages} {605}
  (\bibinfo {year} {1995})}\BibitemShut {NoStop}%
\bibitem [{\citenamefont {Brody}\ and\ \citenamefont
  {Rivier}(1995)}]{Brody1995}%
  \BibitemOpen
  \bibfield  {author} {\bibinfo {author} {\bibfnamefont {D.}~\bibnamefont
  {Brody}}\ and\ \bibinfo {author} {\bibfnamefont {N.}~\bibnamefont {Rivier}},\
  }\href@noop {} {\bibfield  {journal} {\bibinfo  {journal} {Phys. Rev. E}\
  }\textbf {\bibinfo {volume} {51}},\ \bibinfo {pages} {1006} (\bibinfo {year}
  {1995})}\BibitemShut {NoStop}%
\bibitem [{\citenamefont {Dolan}(1998)}]{Dolan1998}%
  \BibitemOpen
  \bibfield  {author} {\bibinfo {author} {\bibfnamefont {B.~P.}\ \bibnamefont
  {Dolan}},\ }\href@noop {} {\bibfield  {journal} {\bibinfo  {journal} {Proc.
  Roy. Soc. A}\ }\textbf {\bibinfo {volume} {454}},\ \bibinfo {pages} {2655}
  (\bibinfo {year} {1998})}\BibitemShut {NoStop}%
\bibitem [{\citenamefont {Janke}\ \emph {et~al.}(2002)\citenamefont {Janke},
  \citenamefont {Johnston},\ and\ \citenamefont {Malmini}}]{Janke2002}%
  \BibitemOpen
  \bibfield  {author} {\bibinfo {author} {\bibfnamefont {W.}~\bibnamefont
  {Janke}}, \bibinfo {author} {\bibfnamefont {D.~A.}\ \bibnamefont
  {Johnston}},\ and\ \bibinfo {author} {\bibfnamefont {R.~P. K.~C.}\
  \bibnamefont {Malmini}},\ }\href@noop {} {\bibfield  {journal} {\bibinfo
  {journal} {Phys. Rev. E}\ }\textbf {\bibinfo {volume} {66}},\ \bibinfo
  {pages} {056119} (\bibinfo {year} {2002})}\BibitemShut {NoStop}%
\bibitem [{\citenamefont {Janke}\ \emph {et~al.}(2003)\citenamefont {Janke},
  \citenamefont {Johnston},\ and\ \citenamefont {Kenna}}]{Janke2003}%
  \BibitemOpen
  \bibfield  {author} {\bibinfo {author} {\bibfnamefont {W.}~\bibnamefont
  {Janke}}, \bibinfo {author} {\bibfnamefont {D.~A.}\ \bibnamefont
  {Johnston}},\ and\ \bibinfo {author} {\bibfnamefont {R.}~\bibnamefont
  {Kenna}},\ }\href@noop {} {\bibfield  {journal} {\bibinfo  {journal} {Phys.
  Rev. E}\ }\textbf {\bibinfo {volume} {67}},\ \bibinfo {pages} {046106}
  (\bibinfo {year} {2003})}\BibitemShut {NoStop}%
\bibitem [{\citenamefont {Brody}\ and\ \citenamefont {Ritz}(2003)}]{Brody2003}%
  \BibitemOpen
  \bibfield  {author} {\bibinfo {author} {\bibfnamefont {D.~C.}\ \bibnamefont
  {Brody}}\ and\ \bibinfo {author} {\bibfnamefont {A.}~\bibnamefont {Ritz}},\
  }\href@noop {} {\bibfield  {journal} {\bibinfo  {journal} {Journal of
  Geometry and Physics}\ }\textbf {\bibinfo {volume} {47}},\ \bibinfo {pages}
  {207} (\bibinfo {year} {2003})}\BibitemShut {NoStop}%
\bibitem [{\citenamefont {You}\ \emph {et~al.}(2007)\citenamefont {You},
  \citenamefont {Li},\ and\ \citenamefont {Gu}}]{You2007}%
  \BibitemOpen
  \bibfield  {author} {\bibinfo {author} {\bibfnamefont {W.-L.}\ \bibnamefont
  {You}}, \bibinfo {author} {\bibfnamefont {Y.-W.}\ \bibnamefont {Li}},\ and\
  \bibinfo {author} {\bibfnamefont {S.-J.}\ \bibnamefont {Gu}},\ }\href@noop {}
  {\bibfield  {journal} {\bibinfo  {journal} {Phys. Rev. E}\ }\textbf {\bibinfo
  {volume} {76}},\ \bibinfo {pages} {022101} (\bibinfo {year}
  {2007})}\BibitemShut {NoStop}%
\bibitem [{\citenamefont {Zanardi}\ \emph {et~al.}(2007)\citenamefont
  {Zanardi}, \citenamefont {Campos~Venuti},\ and\ \citenamefont
  {Giorda}}]{Zanardi2007}%
  \BibitemOpen
  \bibfield  {author} {\bibinfo {author} {\bibfnamefont {P.}~\bibnamefont
  {Zanardi}}, \bibinfo {author} {\bibfnamefont {L.}~\bibnamefont
  {Campos~Venuti}},\ and\ \bibinfo {author} {\bibfnamefont {P.}~\bibnamefont
  {Giorda}},\ }\href@noop {} {\bibfield  {journal} {\bibinfo  {journal} {Phys.
  Rev. A}\ }\textbf {\bibinfo {volume} {76}},\ \bibinfo {pages} {062318}
  (\bibinfo {year} {2007})}\BibitemShut {NoStop}%
\bibitem [{\citenamefont {Zanardi}\ \emph {et~al.}(2008)\citenamefont
  {Zanardi}, \citenamefont {Paris},\ and\ \citenamefont
  {Campos~Venuti}}]{Zanardi2008}%
  \BibitemOpen
  \bibfield  {author} {\bibinfo {author} {\bibfnamefont {P.}~\bibnamefont
  {Zanardi}}, \bibinfo {author} {\bibfnamefont {M.~G.~A.}\ \bibnamefont
  {Paris}},\ and\ \bibinfo {author} {\bibfnamefont {L.}~\bibnamefont
  {Campos~Venuti}},\ }\href@noop {} {\bibfield  {journal} {\bibinfo  {journal}
  {Phys. Rev. A}\ }\textbf {\bibinfo {volume} {78}},\ \bibinfo {pages} {042105}
  (\bibinfo {year} {2008})}\BibitemShut {NoStop}%
\bibitem [{\citenamefont {Paunkovi\ifmmode~\acute{c}\else \'{c}\fi{}}\ \emph
  {et~al.}(2008)\citenamefont {Paunkovi\ifmmode~\acute{c}\else \'{c}\fi{}},
  \citenamefont {Sacramento}, \citenamefont {Nogueira}, \citenamefont
  {Vieira},\ and\ \citenamefont {Dugaev}}]{Paunkovic2008}%
  \BibitemOpen
  \bibfield  {author} {\bibinfo {author} {\bibfnamefont {N.}~\bibnamefont
  {Paunkovi\ifmmode~\acute{c}\else \'{c}\fi{}}}, \bibinfo {author}
  {\bibfnamefont {P.~D.}\ \bibnamefont {Sacramento}}, \bibinfo {author}
  {\bibfnamefont {P.}~\bibnamefont {Nogueira}}, \bibinfo {author}
  {\bibfnamefont {V.~R.}\ \bibnamefont {Vieira}},\ and\ \bibinfo {author}
  {\bibfnamefont {V.~K.}\ \bibnamefont {Dugaev}},\ }\href@noop {} {\bibfield
  {journal} {\bibinfo  {journal} {Phys. Rev. A}\ }\textbf {\bibinfo {volume}
  {77}},\ \bibinfo {pages} {052302} (\bibinfo {year} {2008})}\BibitemShut
  {NoStop}%
\bibitem [{\citenamefont {Quan}\ and\ \citenamefont
  {Cucchietti}(2009)}]{Quan2009}%
  \BibitemOpen
  \bibfield  {author} {\bibinfo {author} {\bibfnamefont {H.~T.}\ \bibnamefont
  {Quan}}\ and\ \bibinfo {author} {\bibfnamefont {F.~M.}\ \bibnamefont
  {Cucchietti}},\ }\href@noop {} {\bibfield  {journal} {\bibinfo  {journal}
  {Phys. Rev. E}\ }\textbf {\bibinfo {volume} {79}},\ \bibinfo {pages} {031101}
  (\bibinfo {year} {2009})}\BibitemShut {NoStop}%
\bibitem [{\citenamefont {Gu}(2010)}]{Gu2010}%
  \BibitemOpen
  \bibfield  {author} {\bibinfo {author} {\bibfnamefont {S.-J.}\ \bibnamefont
  {Gu}},\ }\href@noop {} {\bibfield  {journal} {\bibinfo  {journal} {Int. J.
  Mod. Phys. B}\ }\textbf {\bibinfo {volume} {24}},\ \bibinfo {pages} {4371}
  (\bibinfo {year} {2010})}\BibitemShut {NoStop}%
\bibitem [{\citenamefont {Prokopenko}\ \emph {et~al.}(2011)\citenamefont
  {Prokopenko}, \citenamefont {Lizier}, \citenamefont {Obst},\ and\
  \citenamefont {Wang}}]{Prokopenko2011}%
  \BibitemOpen
  \bibfield  {author} {\bibinfo {author} {\bibfnamefont {M.}~\bibnamefont
  {Prokopenko}}, \bibinfo {author} {\bibfnamefont {J.~T.}\ \bibnamefont
  {Lizier}}, \bibinfo {author} {\bibfnamefont {O.}~\bibnamefont {Obst}},\ and\
  \bibinfo {author} {\bibfnamefont {X.~R.}\ \bibnamefont {Wang}},\ }\href@noop
  {} {\bibfield  {journal} {\bibinfo  {journal} {Phys. Rev. E}\ }\textbf
  {\bibinfo {volume} {84}},\ \bibinfo {pages} {041116} (\bibinfo {year}
  {2011})}\BibitemShut {NoStop}%
\bibitem [{\citenamefont {Marzolino}\ and\ \citenamefont
  {Braun}(2013)}]{Marzolino2013}%
  \BibitemOpen
  \bibfield  {author} {\bibinfo {author} {\bibfnamefont {U.}~\bibnamefont
  {Marzolino}}\ and\ \bibinfo {author} {\bibfnamefont {D.}~\bibnamefont
  {Braun}},\ }\href@noop {} {\bibfield  {journal} {\bibinfo  {journal} {Phys.
  Rev. A}\ }\textbf {\bibinfo {volume} {88}},\ \bibinfo {pages} {063609}
  (\bibinfo {year} {2013})}\BibitemShut {NoStop}%
\bibitem [{\citenamefont {Marzolino}\ and\ \citenamefont
  {Braun}(2015)}]{Marzolino2015}%
  \BibitemOpen
  \bibfield  {author} {\bibinfo {author} {\bibfnamefont {U.}~\bibnamefont
  {Marzolino}}\ and\ \bibinfo {author} {\bibfnamefont {D.}~\bibnamefont
  {Braun}},\ }\href@noop {} {\bibfield  {journal} {\bibinfo  {journal} {Phys.
  Rev. A}\ }\textbf {\bibinfo {volume} {91}},\ \bibinfo {pages} {039902(E)}
  (\bibinfo {year} {2015})}\BibitemShut {NoStop}%
\bibitem [{\citenamefont {Braun}\ \emph {et~al.}(2018)\citenamefont {Braun},
  \citenamefont {Adesso}, \citenamefont {Benatti}, \citenamefont {Floreanini},
  \citenamefont {Marzolino}, \citenamefont {Mitchell},\ and\ \citenamefont
  {Pirandola}}]{Braun2018}%
  \BibitemOpen
  \bibfield  {author} {\bibinfo {author} {\bibfnamefont {D.}~\bibnamefont
  {Braun}}, \bibinfo {author} {\bibfnamefont {G.}~\bibnamefont {Adesso}},
  \bibinfo {author} {\bibfnamefont {F.}~\bibnamefont {Benatti}}, \bibinfo
  {author} {\bibfnamefont {R.}~\bibnamefont {Floreanini}}, \bibinfo {author}
  {\bibfnamefont {U.}~\bibnamefont {Marzolino}}, \bibinfo {author}
  {\bibfnamefont {M.~W.}\ \bibnamefont {Mitchell}},\ and\ \bibinfo {author}
  {\bibfnamefont {S.}~\bibnamefont {Pirandola}},\ }\href@noop {} {\bibfield
  {journal} {\bibinfo  {journal} {Rev. Mod. Phys.}\ }\textbf {\bibinfo {volume}
  {90}},\ \bibinfo {pages} {035006} (\bibinfo {year} {2018})}\BibitemShut
  {NoStop}%
\bibitem [{\citenamefont {Wood}(1992)}]{Wood1992}%
  \BibitemOpen
  \bibfield  {author} {\bibinfo {author} {\bibfnamefont {D.~C.}\ \bibnamefont
  {Wood}},\ }\href@noop {} {\emph {\bibinfo {title} {{The computation of
  polylogarithm}}}},\ \bibinfo {type} {Technical Report 15-92}\ (\bibinfo
  {institution} {University of Kent, Computing Laboratory,30 University of
  Kent, Canterbury, UK},\ \bibinfo {year} {1992})\BibitemShut {NoStop}%
\bibitem [{\citenamefont {van~den Berg}\ \emph {et~al.}(1984)\citenamefont
  {van~den Berg}, \citenamefont {Lewis},\ and\ \citenamefont
  {de~Smedt}}]{vandenBerg1984}%
  \BibitemOpen
  \bibfield  {author} {\bibinfo {author} {\bibfnamefont {M.}~\bibnamefont
  {van~den Berg}}, \bibinfo {author} {\bibfnamefont {J.~T.}\ \bibnamefont
  {Lewis}},\ and\ \bibinfo {author} {\bibfnamefont {P.}~\bibnamefont
  {de~Smedt}},\ }\href@noop {} {\bibfield  {journal} {\bibinfo  {journal} {J.
  Stat. Phys.}\ }\textbf {\bibinfo {volume} {37}},\ \bibinfo {pages} {697}
  (\bibinfo {year} {1984})},\ \bibinfo {note} {notice the typo in equation
  (2.19) of this reference: one $\mu$ in the grand canonical thermodynamic
  potential has been replaced by $p$. However, it is clear from the proof, e.g.
  see equations (4.4,4.9) in the reference, that our equation (91) is the
  correct formula.}\BibitemShut {Stop}%
\end{thebibliography}

%

\end{document}